\documentclass[10pt,conference]{IEEEtran}
\IEEEoverridecommandlockouts
\usepackage{cite}
\usepackage{amsmath,amssymb,amsfonts}
\usepackage{algorithm}
\usepackage
{algpseudocode}
\usepackage{etoolbox}
\usepackage{multirow}
\usepackage[table,xcdraw]{xcolor}
\usepackage{graphicx}
\usepackage{textcomp}
\usepackage{xcolor}
\usepackage{url}
\usepackage{hyperref}
\usepackage{subcaption}

\usepackage{amsthm}
\newtheorem{theorem}{Theorem}

\newcommand{\squishlist}{
   \begin{list}{$\bullet$}
    { \setlength{\itemsep}{2pt}
    \setlength{\parsep}{0pt}
      \setlength{\topsep}{5pt}     \setlength{\partopsep}{0pt}
      \setlength{\leftmargin}{1.35em} \setlength{\labelwidth}{1em}
      \setlength{\labelsep}{0.5em} } }

\newcommand{\squishlisttwo}{
   \begin{list}{$\bullet$}
    { \setlength{\itemsep}{0pt}    \setlength{\parsep}{0pt}
      \setlength{\topsep}{0pt}     \setlength{\partopsep}{0pt}
      \setlength{\leftmargin}{1.35em} \setlength{\labelwidth}{1em}
      \setlength{\labelsep}{0.5em} } }

\newcommand{\squishlistBigLabel}{
   \begin{list}{$\bullet$}
    { \setlength{\itemsep}{2pt}
    \setlength{\parsep}{0pt}
      \setlength{\topsep}{5pt}     \setlength{\partopsep}{0pt}
      \setlength{\leftmargin}{2.35em} \setlength{\labelwidth}{1.8em}
      \setlength{\labelsep}{0.5em} } }

\newcommand{\squishend}{
    \end{list}  }

\makeatletter
\newcommand*{\algrule}[1][\algorithmicindent]{%
  \makebox[#1][l]{%
    \hspace*{.1em}
    \vrule height .75\baselineskip depth .25\baselineskip
  }
}

\newcount\ALG@printindent@tempcnta
\def\ALG@printindent{%
    \ifnum \theALG@nested>0
    \ifx\ALG@text\ALG@x@notext
    \else
    \unskip
    \ALG@printindent@tempcnta=1
    \loop
    \algrule[\csname ALG@ind@\the\ALG@printindent@tempcnta\endcsname]%
    \advance \ALG@printindent@tempcnta 1
    \ifnum \ALG@printindent@tempcnta<\numexpr\theALG@nested+1\relax
    \repeat
    \fi
    \fi
}
\patchcmd{\ALG@doentity}{\noindent\hskip\ALG@tlm}{\ALG@printindent}{}{\errmessage{failed to patch}}
\patchcmd{\ALG@doentity}{\item[]\nointerlineskip}{}{}{} 
\makeatother
\algrenewcommand\algorithmicindent{1.1em}%

\newcommand{\approach}{fPMC}

\begin{document}

\title{Fast Parametric Model Checking through Model Fragmentation}

\makeatletter
\newcommand{\linebreakand}{%
  \end{@IEEEauthorhalign}
  \hfill\mbox{}\par
  \mbox{}\hfill\begin{@IEEEauthorhalign}
}
\makeatother

\author{\IEEEauthorblockN {Xinwei Fang}
\IEEEauthorblockA{\textit{Department of Computer Science}\\
\textit{University of York, UK}\\
xinwei.fang@york.ac.uk}
\and
\IEEEauthorblockN {Radu Calinescu}
\IEEEauthorblockA{\textit{Department of Computer Science}\\
\textit{University of York, UK}\\
radu.calinescu@york.ac.uk}\\
\linebreakand
\IEEEauthorblockN {Simos Gerasimou}
\IEEEauthorblockA{\textit{Department of Computer Science }\\
\textit{University of York, UK}\\
simos.gerasimou@york.ac.uk}
\and 
\IEEEauthorblockN {Faisal Alhwikem}
\IEEEauthorblockA{\textit{Department of Computer Science}\\
\textit{University of York, UK}\\
faisal.alhwikem@york.ac.uk}
}

\maketitle

\begin{abstract}
Parametric model checking (PMC) computes algebraic formulae that express key non-functional properties of a system (reliability, performance, etc.) as rational  functions of the system and environment parameters. In software engineering, PMC formulae can be used during design, e.g., to analyse the sensitivity of different system architectures to parametric variability, or to find optimal system configurations. They can also be used at runtime, e.g., to check if non-functional requirements are still satisfied after environmental changes, or to select new configurations after such changes. However, current PMC techniques do not scale well to systems with complex behaviour and more than a few parameters. Our paper introduces a fast PMC (fPMC) approach that overcomes this limitation, extending the applicability of PMC to a broader class of systems than previously possible. To this end, fPMC partitions the Markov models that PMC operates with into \emph{fragments} whose reachability properties are analysed independently, and obtains PMC reachability formulae by combining the results of these fragment analyses. To demonstrate the effectiveness of fPMC, we show how our fPMC tool can analyse three systems (taken from the research literature, and belonging to different application domains) with which current PMC techniques and tools struggle.
\end{abstract}


\section{Introduction}\label{sec:intro}

Parametric model checking (PMC)~\cite{Daws:2004:SPM:2102873.2102899,Hahn2011,Jansen2014,RaduePMC} is a formal technique for analysing the reliability and performance of systems with stochastic behaviour. The underlying concepts are very simple. Consider a foreign exchange trading (FX) system that obtains up-to-date currency exchange rates from two external web services, by first invoking one of the services, and only invoking the other service if the first service is busy (i.e., if its invocation times out). If the probabilities that the two services are not busy are $p_1$ and $p_2$, then the system will successfully obtain the exchange rates with probability $p_\mathsf{succ}=p_1+(1-p_1)p2$. 

Expressing the non-functional properties of software systems as \emph{closed-form formulae}\footnote{i.e., mathematical expressions containing constants, variables and simple binary operations ($+$, $-$, $\times$, $/$, etc.) that can be computed in constant time} like this has numerous benefits. At design time, the PMC formulae can be evaluated very efficiently to compare architectures associated with different parameter values, e.g., in software product lines~\cite{ghezzi2013model,DBLP:conf/splc/GhezziS11}. At runtime, they can be used to efficiently re-verify the satisfaction of non-functional requirements after environmental parameter changes~\cite{10.1145/2884781.2884814}, and to select new optimal values for the configuration parameters~\cite{Filieri2011,Filieri2013}. They also enable the analysis of the non-functional property sensitivity to variations in the system parameters~\cite{DBLP:journals/tse/FilieriTG16}, and the computation of confidence intervals for the analysed non-functional properties~\cite{calinescu2016formal,calinescu2016fact}.
For instance, if the FX system from our earlier example must operate with  $p_\mathsf{succ}\geq 0.99$ and its developers know (e.g., from service-level agreements) that $p_1\geq 0.95$, they can compute the minimum acceptable success probability for the second web service as $p_2=(0.99-0.95)/(1-0.95)=0.8$. As another illustration, if unit testing is used to establish $[0.8,0.9]$ as a $.95$ confidence interval for both $p_1$ and $p_2$ from our FX example, then a $.95^2\approx .9$ confidence interval for $p_\mathsf{succ}$ is given by $[\min_{p_1,p_2\in [0.8,0.9]}p_\mathsf{succ},\max_{p_1,p_2\in [0.8,0.9]}p_\mathsf{succ}]=[0.96,0.99]$.

Of course, for non-trivial systems neither the PMC calculations nor the formulae they produce are this simple. In fact, they both become so complex that, despite significant advances in recent years, current PMC methods~\cite{Daws:2004:SPM:2102873.2102899,Hahn2011,Jansen2014} and model checkers~\cite{param,prism,storm} do not scale well (i.e., time out, run out of memory, or yield formulae that are too large to evaluate) for many realistic systems with more than a few parameters. 

\begin{figure}
	\centering
    \includegraphics[width=\hsize]{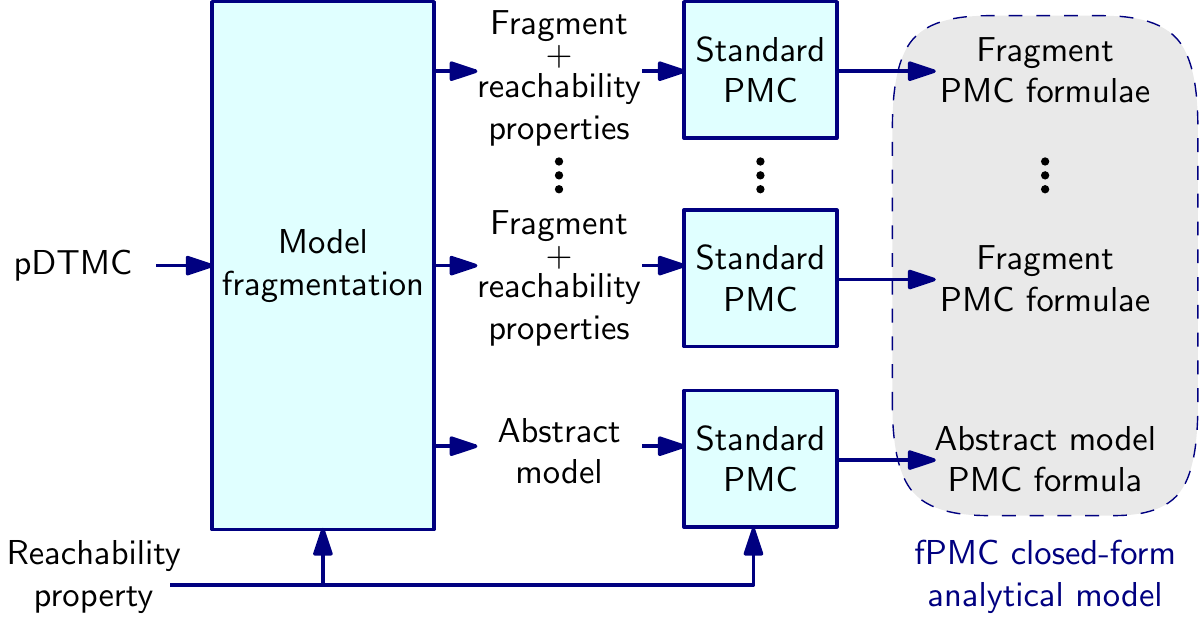}	
    \caption{Fast parametric model checking process}
    \label{fig:approach}
    
    \vspace*{-2mm}
\end{figure}

Our paper introduces a fast PMC (fPMC) technique that complements the advances from~\cite{Daws:2004:SPM:2102873.2102899,Hahn2011,Jansen2014}, enabling the efficient parametric model checking of reachability properties for parametric discrete-time Markov chains (pDTMCs) of systems with more complex behaviour and larger numbers of parameters than previously possible. fPMC \emph{extends} and \emph{automates} a recently proposed theoretical framework for efficient parametric model checking~\cite{RaduePMC}.

As shown in Fig.~\ref{fig:approach}, fPMC uses a divide-and-conquer strategy to partition a pDTMC under analysis into \emph{fragments} (i.e., subsets of model states and transitions that satisfy well-defined rules detailed later in the paper) accompanied by fragment-level reachability properties, and derives an \emph{abstract model} for the analysed pDTMC by replacing each of its fragments with a single state. The fPMC fragments and abstract model are pDTMCs in their own right, and can be analysed individually (using standard PMC techniques~\cite{Daws:2004:SPM:2102873.2102899,Hahn2011,Jansen2014}) to produce a set of PMC formulae whose combination provides a \emph{closed-form analytical model} for the system-level property of interest.

The main contributions of our paper are: 
\begin{enumerate}
\item The first \emph{model fragmentation} algorithm for the automated partition of pDTMCs into fragments with the characteristics defined by our theoretical framework from~\cite{RaduePMC}, which shows how model fragments can be exploited, but does \emph{not} provide a generally applicable model fragmentation technique. 

\item New theoretical results that allow the formation of model fragments from subsets of pDTMC states that do not (initially) meet the definition of a fragment. The new results define valid pDTMC structural modifications that enable the formation of model fragments of the right size for their ``standard'' PMC analysis to be feasible.

\item A prototype fPMC tool that implements the new model fragmentation algorithm and theoretical results introduced in the paper. The fPMC tool invokes the Storm model checker~\cite{storm} for the PMC of model fragments, fully automating the synthesis of closed-form analytical models for the analysed reachability properties. 

\item An extensive evaluation that demonstrates the fPMC effectiveness at analysing pDTMC reachability properties for systems from three application domains.
\end{enumerate}

We organised the rest of the paper as follows.  Section~\ref{sec:background} provides the required background on parametric model checking. Section~\ref{sec:example} introduces a running example that we use to illustrate the use of our fPMC technique, which are detailed in Section~\ref{sec:approach}. Sections~\ref{sec:implementation} and~\ref{sec:evaluation} describe our fPMC implementation and evaluation, respectively. Finally, Section~\ref{sec:relatedWork} discusses related work, and Section~\ref{sec:conclusions} summarises our results and suggests directions for future research.

\section{Preliminaries}\label{sec:background}

\subsection{Parametric Model Checking}
\label{pmc}

Parametric model checking (PMC) is a formal technique for the symbolic analysis of Markov chains whose transition probabilities are specified as rational functions over a set of continuous variables~\cite{Daws:2004:SPM:2102873.2102899,Jansen2014,Hahn2011}.
Formally, a parametric discrete-time Markov chain is a tuple $ D=(S,s_0,\textbf{P}, L),$ where: $S$ is a finite set of states; $s_0 \in S$ is the initial state;\mbox{$\textbf{P}: S \times S \rightarrow [0,1]$} is a transition probability matrix such that for any states \mbox{$s, s' \in S$}, $\textbf{P}(s,s')$ gives the probability of transitioning from $s$ to $s'$, and, for any $s \in S$,  $\sum_{s' \in S}, \textbf{P}(s,s')=1$; $L: S \rightarrow 2^\mathit{AP}$ is a labelling function that maps every state $s \in S$ to elements of a set of atomic propositions $\mathit{AP}$ that hold in that state. A state $s \in S$ is \emph{absorbing} if $\textbf{P}(s,s)=1$ and $\textbf{P}(s,s')=0$ for all $s \ne s'$, and a \emph{transient} state otherwise.

When used in software performance and reliability engineering, 
the pDTMC states map to relevant configurations of the modelled system; and pDTMC transitions capture possible transitions between those states, corresponding to the feasible changes between their associated configurations. PMC is supported by the probabilistic model checkers  PARAM~\cite{param}, PRISM~\cite{prism} and Storm~\cite{storm}. 
These tools compute \emph{closed-form formulae} for pDTMC properties 
specified in probabilistic computation tree logic (PCTL)~\cite{hansson1994logic,bianco_alfaro_1995,Ciesinski04G} and defined (as in probabilistic model checking~\cite{baier2008principles,DBLP:conf/vmcai/Katoen10,KNP07a,Kwiatkowska2018,KNP10c}) by the grammar: 
\begin{equation}
\label{eq:pctl}
\begin{array}{l}
    \Phi::=  true \,\vert\, a \,\vert\, \Phi \wedge \Phi \,\vert\, \neg \Phi \,\vert\, \mathcal{P}_{\!=?} [\Psi] \\
    \Psi::= X \Phi \;\vert\; \Phi\; \mathrm{U}\; \Phi \;\vert\; \Phi\; \mathrm{U}^{\leq k}\, \Phi\\
\end{array}
\end{equation}
where $\Phi$  is a \emph{state formula} and $\Psi$ is a \emph{path formula}, $k \in \mathbb{N}_{>0}$ is a timestep bound and $\alpha \in AP$ is an atomic proposition.

The PCTL semantics is defined using a satisfaction \mbox{relation $\models$} over the states $S$. Given a state $s$ of a Markov chain $D$, $s\models \Phi$ means ``$\Phi$ holds in state $s$'', and we have: always $s\models true$; $s \models a$ iff $a\in L(s)$; $s \models \neg \Phi$ iff $\neg (s\models \Phi)$; and $s\models \Phi_1 \wedge \Phi_2$ iff $s\models \Phi_1$ and $s\models \Phi_2$. The \emph{time-bounded until formula} $\Phi_1\, \mathrm{U}^{\leq k}\, \Phi_2$ holds for a path iff $\Phi_1$ holds in the first $i<k$ path states and $\Phi_2$ holds in the $(i+1)$-th path state; and the \emph{unbounded until formula} $\Phi_1\,\mathrm{U}\, \Phi_2$ removes the bound $k$ from the  time-bounded until formula. 
The \emph{next formula} $X \Phi$ holds if $\Phi$ is satisfied in the next state. 
The state formula $\mathcal{P}_{\!=?} [\Psi]$ specifies the probability  that paths starting at a chosen state $s$ satisfy a path property $\Psi$. \emph{Reachability properties}  $\mathcal{P}_{\!=?} [\mathsf{true}\, \mathrm{U}\, \Phi]$ are equivalently written as $\mathcal{P}_{\!=?} [\mathsf{F}\, \Phi]$ or $\mathcal{P}_{\!=?} [\mathsf{F}\, R]$, where $R\!\subseteq\! S$ is the set of  states in which $\Phi$ holds. For a full description of the PCTL semantics, see~\cite{bianco_alfaro_1995,hansson1994logic}.

\subsection{Parametric Model Checking Using Model Fragments}\label{subsec:fragmentationTheory}

We summarise the concept of a pDTMC fragment and its exploitation by the PMC approach introduced in~\cite{RaduePMC}, where  the results from this section are taken from. As shown in Fig.~\ref{fig:fragment-diagram}, a fragment of a pDTMC $D=(S,s_0,\textbf{P}, L)$ is a tuple 
\begin{equation}
    \label{eq:fragment}
  F=(Z,z_0,Z_\mathsf{out}),
\end{equation}
where:
$Z\subset S$ is a subset of transient MC states; 
$z_0$ is the (only) \emph{input state} of $F$, i.e., $\{z_0\}\!=\!\{z\!\in\! Z\mid \exists s\!\in\! S\!\setminus\! Z\:.\: \mathbf{P}(s,z)>0\}$;
$Z_\mathsf{out} =\{z\in Z 
\mid \exists s\in S\setminus Z\:.\: \mathbf{P}(z,s)>0\}$ is the non-empty set of \emph{output states} of $F$, and all outgoing transitions from the output states are to states outside $Z\setminus\{z_0\}$, i.e., $\mathbf{P}(z,z')=0$ for all $(z,z')\in Z_\mathsf{out}\times (Z\setminus\{z_0\})$.

\begin{figure}
	\centering
    \includegraphics[width=\linewidth]{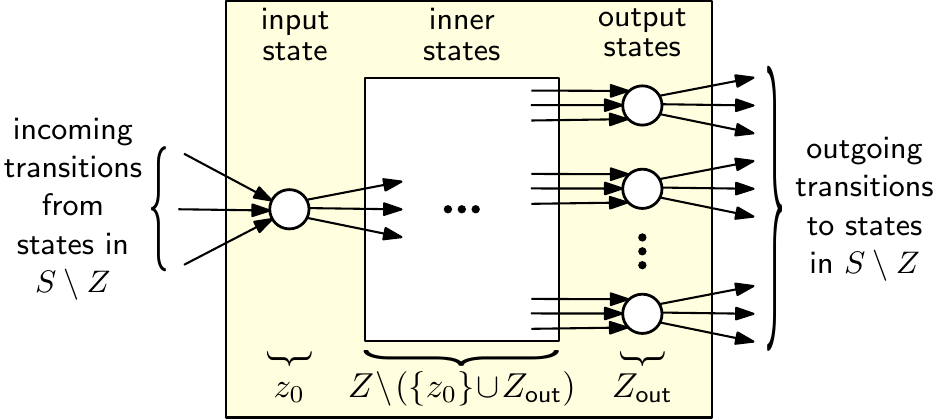}	
    \caption{Fragment of a pDTMC model}
    \label{fig:fragment-diagram}
    
    \vspace*{-2mm}
\end{figure}

Fragments are not strongly connected components (SCCs) of the \emph{graph induced by the pDTMC}\footnote{i.e., the directed graph comprising a vertex for each pDMTC state and an edge between each pair of vertices that correspond to pDTMC states between which a transition is possible} states and transitions. For instance, the ``inner states'' region of the fragment from Fig.~\ref{fig:fragment-diagram} can contain one or more SCCs, and an outgoing transition from an output state in $Z_\mathsf{out}$ can reach a state $s$ comprising an outgoing transition back to the input state $z_0$. Any subset of model states satisfying the constraints described earlier in this section form a fragment. However, partitioning a pDTMC or a part of a pDTMC into fragments is not always possible. For example, fully connected subsets of states cannot be split into fragments because none of their states can be an inner fragment state (as shown in Fig.~\ref{fig:fragment-diagram}) as it will have incoming transitions from outside states and outgoing transitions to outside states; restructuring using the techniques described later in Section~\ref{ssec:modelRestructuring} is not possible either, as the premises for these techniques are violated. 

The theoretical framework from~\cite{RaduePMC} shows that, if a fragment $F$ of a pDTMC $D$ is identified, then the ``monolithic'', one-step PMC of reachability, unbounded until and (not covered in our paper) reward properties~\cite{andova2003discrete} of $D$ can be replaced by the following equivalent steps:
\begin{enumerate}
\item Compute PMC formulae for the probabilities of reaching all the output states of fragment $F$ starting from its input state $z_0$; this amounts to the PMC of $\# Z_\mathsf{out}$ reachability properties of a pDTMC built from $F$ removing the incoming transitions of $z_0$ and by replacing the outgoing transitions of each output state in $Z_\mathsf{out}$ with a ``self-loop'' transition of probability $1$.
\item Build an \emph{abstract pDTMC model} $D'$ from $D$ by replacing the states in $Z$ and their internal transitions with a single state $z'$ whose incoming transitions (and transition probabilities) are those of $z_0$. Additionally, state $z'$ has outgoing transitions to every state that one or more states from $Z_\mathsf{out}$ have outgoing transitions to in $D$; and the probabilities of these transitions can be expressed in terms of the reachability properties computed in step~1. Due to space constraints, we do not provide the expressions for these transition probabilities in the paper; the interested reader can find them in~\cite{RaduePMC}.
\item Compute the PMC formula for the original property under analysis, for the abstract model from step~2.
\item Combine the PMC from step~1 and the PMC formula from step~3 into a system of equations.
\end{enumerate}
The system of equations from step~4 provides a closed-form analytical model for the analysed property. This analytical model is equivalent to the PMC formula obtained by analysing the original pDTMC in one step. 

Because the PMC analyses from steps~1 and~3 are performed on models that are smaller and simpler than $D$, this four-step PMC approach is often faster, produces much simpler closed-form expressions, or succeeds for more complex models than monolithic PMC can handle. However, no method for partitioning pDTMCs into fragments is proposed in~\cite{RaduePMC}. Instead, the approach is only exploited for specific types of component-based systems (i.e., service-based systems and multi-tier architectures) for which: (i)~pDTMCs can be obtained by combining manually pre-built pDTMC models of their components; and (ii)~the component pDTMCs form fragments in the system pDTMC. This limits the applicability of the approach (to specific types of systems for which new pDTMCs are assembled following this recipe), requires additional effort and expertise, and can be error prone due to the manual steps involved.

\section{Running Example}\label{sec:example}
We illustrate our \approach\ approach using a six-operation service-based system from the area of foreign exchange trading (FX) introduced in~\cite{Gerasimou2015:ASE,gerasimou2018synthesis}. The FX system implements the workflow shown in Fig.~\ref{fig:fxworkflow} and described briefly below.

\vspace*{1mm}\noindent
\textbf{FX description.}
A trader can use FX in two execution modes. In the \textit{expert} mode, FX runs a loop that analyses market activity, identifies patterns that satisfy the trader's objectives, and automatically carries out trades. Thus, the \emph{Market watch} operation extracts real-time exchange rates (bid/ask price) of selected currency pairs. This data is used by a \emph{Technical analysis} operation that evaluates the current trading conditions, predicts future price movement, and decides if the trader's objectives are: (i)~``met'' (causing the invocation of an \emph{Order} service to carry out a trade); (ii)~``not met'' (resulting in a new \emph{Market watch} invocation); or (iii)~an error occured (triggering an \emph{Alarm} operation to notify the trader about discrepancies/opportunities not covered by the trading objectives). In the \emph{normal} mode, FX assesses the economic outlook of a country using a \emph{Fundamental analysis} operation that collects, analyses and evaluates information (e.g., news reports, economic data and political events), and provides an assessment on the country's currency. If satisfied with this assessment, the trader can use the \emph{Order} operation to sell/buy currency; then, a \emph{Notification} operation confirms the completion of the trade. 

\begin{figure}
	\centering
    \includegraphics[width=\linewidth]{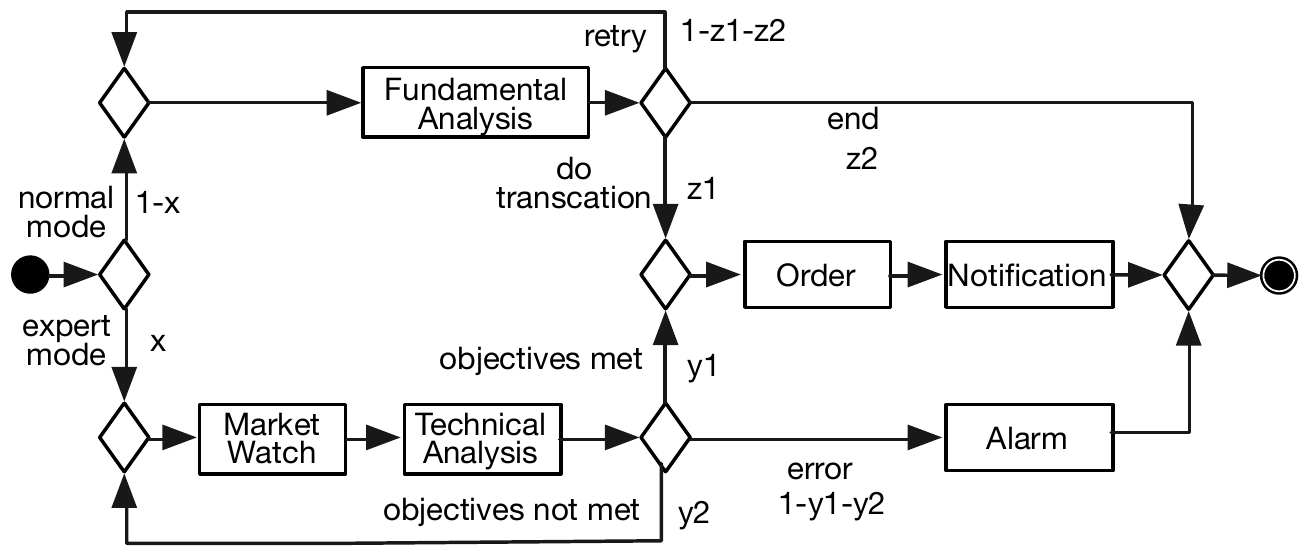}	
    \caption{FX service-based system, where $x$, $y1$, $y2$, $z1$, $z2$ are the (unknown) probabilities of different execution paths}
    \label{fig:fxworkflow}
    
    \vspace*{-2mm}
\end{figure}

Given its business-critical nature, assume that the software architect aims at designing an FX system with high reliability. Thus, for the $i$-th operation FX uses two functionally-equivalent service implementations and adopts a \emph{sequential execution strategy with retry (SEQ\_R)} based on which the two services per operation are invoked in order.
If the first service fails, it is re-invoked with probability $r_{i1}$, whereas the operation is attempted using the second service with probability $1-r_{i1}$.
If the execution of the second service fails, it is retried with probability $r_{i2}$; otherwise, with probability $1-r_{i2}$, the entire system execution fails. 
Finally, assume that the selected service implementations per FX operation is based on the analysis results of the system-level probability of successfully completing the handling of a request.

\begin{figure}
	\centering
	\includegraphics[width=\linewidth]{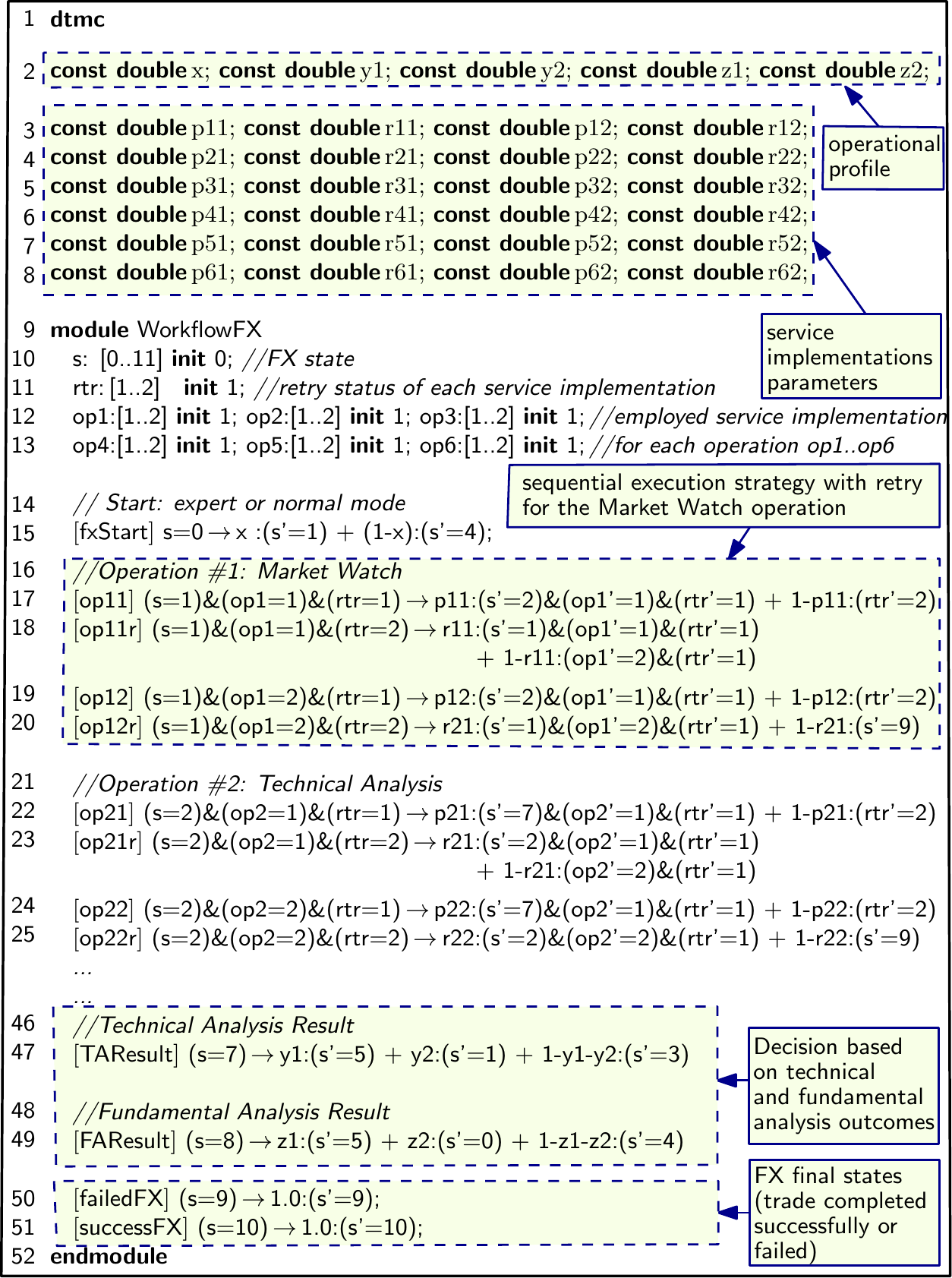}
 		\caption{pDTMC model of the FX system.}
    \label{fig:fxmodel}
    
    \vspace*{-2mm}
\end{figure}

\vspace*{1mm}\noindent
\textbf{FX pDTMC.}
Fig.~\ref{fig:fxmodel} shows the pDTMC of the FX system specified in the modelling language of the PRISM model checker~\cite{prism}.
The model comprises a \emph{WorkflowFX} module modelling the FX workflow (lines 9--52), and the parameters associated with the operational profile (line 2) and with the service implementations for each FX operation (lines 3--8); $p_{ij}$ and $r_{ij}$ signify the probability of successful execution and the probability of retrying the $i$-th operation using the $j$-th service implementation, respectively. The model uses parameters both for the operational profile of the system and for its configurable aspects because the former is typically unknown when the model is produced, and the latter enables the exploration and selection of good configuration parameter values.

Within the \emph{WorkflowFX} module, the local variable \emph{s} (line 10) models the state of the FX system, while $op_i$ (line 12--13) and \emph{retry} (line 11) encode the employed service implementation per operation and the retry status of a service implementation, respectively. 
Following the selection of the expert FX mode (line 15), the \emph{Market watch} operation with retry and two service implementations is executed (lines 16--20). The first service implementation succeeds with probability $p_{11}$ and FX moves to the \emph{Technical analysis} operation, fails with probability $1-p_{11}$ and retries with probability $r_{11}$ (lines 17--18); otherwise, the second service implementation is executed and succeeds or is re-invoked with probabilities $p_{12}$ and $r_{12}$, respectively. If both service implementations fail, the FX execution terminates (line 50). The other FX operations function similarly  (lines 21--45) but we omit their details due to space constraints.
The pDTMC model of the FX system produces the directed graph shown in Fig.~\ref{fig:fxExample}, which comprises 29 states and 58 transitions and with the initial, failed and successful states coloured in blue, red and green, respectively.

Given this pDTMC model, we used PRISM~\cite{prism} and Storm~\cite{storm} to obtain the closed-form PMC formula for the probability of successfully handing a request (i.e., to reach the succeed state) encoded in PCTL as  $\mathcal{P}_{=?}[\mathrm{F}\; s\!=\!10]$.
Despite the small number of states and transitions, neither model checker was able to compute the formula within an hour (on a computer with the specification from Section~\ref{ssec:experimentalSetup}). 
We explain next how \approach\ supports the computation of those formulae through 
automated model fragmentation.

\begin{figure}
	\centering
	\includegraphics[width=\linewidth]{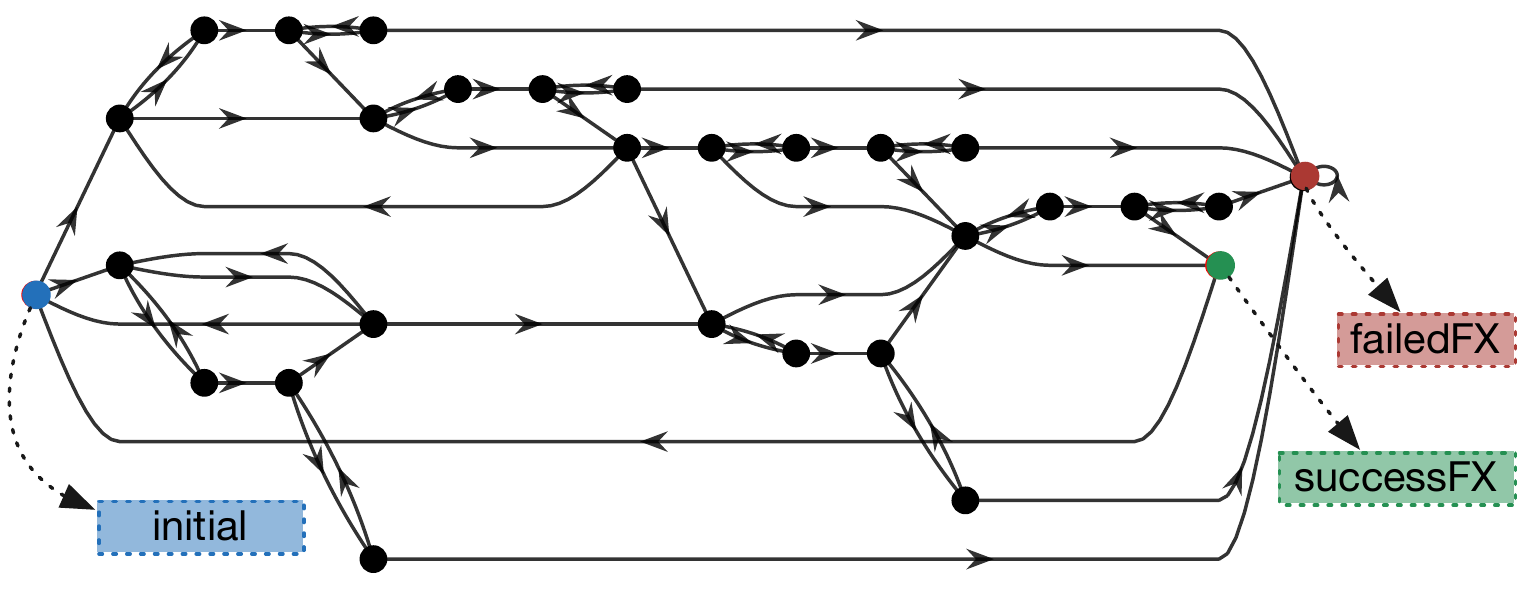}
		\caption{The directed graph induced by the FX pDTMC from Fig.~\ref{fig:fxmodel}, 
		showing the initial, failedFX and successFX states.}
    \label{fig:fxExample}
    
    \vspace*{-2mm}
\end{figure}

\section{fPMC Approach}\label{sec:approach}

\subsection{Markov Chain Fragmentation Algorithm \label{ssec:algorithm}}

The \approach\ partition of a pDTMC into fragments is performed by function \textsc{Fragmentation} from Algorithm~\ref{algorithm:fragmentation}. This function takes three arguments: 
\squishlist
\item the graph $\mathcal{G}(V,E)$ induced by the pDTMC; 
\item the set of states $R$ whose reachability $\mathcal{P}_{=?}[\mathsf{F}\; R]$ is being analysed; 
\item a \emph{threshold} parameter $\alpha \geq 1$ whose role is described below,
\squishend
and returns a set of pDTMC fragments $\mathit{FS}$, i.e., a set of tuples with the structure from Equation~\eqref{eq:fragment}.
As not all pDTMC states can be included into a fragment that follows the definition from Section~\ref{subsec:fragmentationTheory} (e.g., states whose only outgoing transition is a self-loop transition do not satisfy the constraints associated with any type of fragment state), we extend this definition to include single-state fragments $F\!=\!(\{s\},s,\{s\})$ for any pDTMC state $s$.

The algorithm starts by placing into $\mathit{FS}$ single-state fragments for each state in the reachability set $R$ (line~\ref{l:satisfy}). We preserve the elements of $R$ as single-state fragments so they end up as states in the \approach\ abstract model (see Fig.~1 and the description from Section~\ref{subsec:fragmentationTheory}), so the reachability property $\mathcal{P}_{=?}[\mathsf{F}\; R]$ (which refers to these states) can be analysed for the abstract model. Next, the algorithm generates additional fragments in each iteration of the for loop from lines~\ref{l:forS}--\ref{l:forE} as follows. First, a node $z_0$ not yet included in any fragment is selected\footnote{The selection order for $z_0$ has an impact on the fragmentation of the pDTMC, and thus on its parametric model checking time; however, finding the best selection order is outside the scope of this paper.} (line~\ref{l:forS}) and inserted into the fragment state set $Z$, while the fragment output set $Z_\mathsf{out}$ is initialised to the empty set (line~\ref{l:startFragment}).  A stack $T$ is then populated with the states reached by outgoing transitions from $z_0$ by invoking (in line~\ref{l:grow1}) the function \textsc{Traverse} from Algorithm~\ref{algorithm:fragmentgrow}. Each state $w$ from this stack is processed by the while loop from lines~\ref{l:whileS}--\ref{l:whileE}, ending up in $Z_\mathsf{out}$ if it satisfies the constraints associated with output fragment states (lines~\ref{l:ifOutput1} and~\ref{l:ifOutput2}). When $w$ does not satisfy these constraints, two options are possible (lines~\ref{l:check-alphaS}--\ref{l:check-alphaE}):

\squishlist
\item If $Z$ has accumulated fewer states than the threshold $\alpha$, the graph traversal function \textsc{Traverse} is invoked again to add to the stack the predecessor and successor vertices of $w$ that are not already in the fragment (line~\ref{l:grow2}). 
\item Otherwise, a restructuring of the graph (see Section~\ref{ssec:modelRestructuring}) is invoked (line~\ref{l:restructure}) to ``force'' $w$ into becoming an output state, and thus to enable the formation of a fragment. 
\squishend
In this way, the threshold $\alpha$ provides a soft upper bound for the fragment size, deciding whether fPMC will continue model traversal (adding states into the stack) or attempt model restructuring by forcing the currently analysed state to become an output state. As the result, a larger $\alpha$ is expected to lead to larger fragments. When this bound, $\alpha$, is reached, the model restructuring techniques detailed in Section~\ref{ssec:modelRestructuring} are used to force the formation of a valid fragment. 
The fragment candidate $(Z,z_0,Z_\mathsf{out})$ finalised by the while loop is then ``downgraded'' to a single-state fragment if it does not meet the definition from Section~\ref{subsec:fragmentationTheory} (lines~\ref{l:validfragmentS}--\ref{l:validfragmentE}), after which it is added to the fragment set $\mathit{FS}$ (line~\ref{l:addFragment}). Additionally, the states $Z$ of this fragment are added to the set $R$ of states already assigned to fragments (line~\ref{l:addToR1}), ensuring that the loop starting in line~\ref{l:forS} does not reuse them to form other fragments.

Finally, the $\textsc{Traverse}$ function from  Algorithm~\ref{algorithm:fragmentgrow} takes a vertex $w$ and examines its incoming edges (if $w$ is not the initial fragment state $z_0$,  lines~\ref{l:ifInputS}--\ref{l:ifInputE}), and its outgoing edges (at all times, lines~\ref{l:OutputS}--\ref{l:OutputE}). Vertices connected to $w$ by these incoming and outgoing edges and not already in the set of fragment states $Z$ are collected into the input vertex set $I$ (line~\ref{l:inputs}) and output vertex set $O$ (line~\ref{l:OutputS}), respectively. The former are then added to the stack $T$ if none of them belongs to an existing fragment (lines~\ref{l:wAddedIf} and~\ref{l:wAdded}). Otherwise, the non-input vertex $w$ has incoming edge(s) from another fragment, violating the condition that incoming transitions to inner and output fragment states are only allowed from states within the same fragment; hence, $w$ becomes a single state fragment and the graph traversal is terminated (lines~\ref{l:wFragment2S}--\ref{l:wFragment2E}). Lastly, if the output set $O$ has at least a vertex not in other fragments (line~\ref{l:notSubsetOfR}), growing the fragment under construction with $w$'s successors may be feasible, and thus the for loop in lines~\ref{l:OutputMid}--\ref{l:OutputE} places the $O$ vertices not in other fragments on the stack $T$ (line~\ref{l:pushO}) and attempts to restructure the graph to allow fragment formation if these vertices are already part of other fragments (line~\ref{l:restructure2}). 

\newlength{\textfloatsepsave} 
\setlength{\textfloatsepsave}{\textfloatsep} 
\setlength{\textfloatsep}{8pt} 

\begin{algorithm}[t]
	\caption{pDTMC model fragmentation 
	}\label{algorithm:fragmentation}
	\begin{small}
		\renewcommand{\baselinestretch}{1}
		\begin{algorithmic}[1]
			\Function{Fragmentation}{$\mathcal{G}(V,E), R, \alpha$}
            
            \State $\mathit{FS} \gets \left\{\left(\{r\}, r,\{r\}\right) \;|\; r\in R\right\}$ 
            \label{l:satisfy}

			\ForAll{$z_0 \in V\setminus R$ } \label{l:forS}
            \State $Z \gets \{ z_0 \}$, $Z_\mathsf{OUT} \gets \{\}$ \label{l:startFragment} 
			    \State $T \gets \textsc{EmptyStack()}$ 
			        \State $\textsc{Traverse}(\mathcal{G},z_0, T,\mathit{FS},R,Z,\mathsf{true})$ \Comment{Alg.~\ref{algorithm:fragmentgrow}}\label{l:grow1}
			
    			\While{$\neg \textsc{Empty}(T)$}\label{l:whileS}
    			    \State $w \gets T.\textsc{Pop}()$ 
    		        \If {$\{i \:|\: (i,\!w)\!\in\! E\}\!\subseteq\! Z \wedge \{o \:|\: (w,\!o)\!\in\! E\} \!\subseteq\! V\!\setminus\!\! Z\!$}  \label{l:ifOutput1} 
    		           \State $Z_\mathsf{OUT} \gets Z_\mathsf{OUT} \cup \{w\}$ \label{l:ifOutput2}
        		        \Else
        			    \If{$\#Z < \alpha$} \label{l:check-alphaS}
    			            \State $\!\textsc{Traverse}(\mathcal{G},w,T,\mathit{FS},R,Z,\mathsf{false})$ \Comment{$\!\!$Alg.~\ref{algorithm:fragmentgrow}}\label{l:grow2}	
    			        \Else
    			            \State $\textsc{Restructure}(\mathcal{G},w)$ \Comment{$\!\!$Section~\ref{ssec:modelRestructuring}}\label{l:restructure}
    		            \EndIf \label{l:check-alphaE}
    	            \EndIf
                    \State $Z \gets Z \cup \{ w \}$ \label{l:extendFragment} 
    			\EndWhile\label{l:whileE} 

            \If{$\neg\textsc{ValidFragment}((Z,z_0,Z_\mathsf{OUT}))$}\label{l:validfragmentS}
                \State $Z \gets \{z_0\}$, $Z_\mathsf{OUT} \gets \{z_0\}$
                \label{l:degenerate2}
            \EndIf \label{l:validfragmentE}
            \State $\mathit{FS} \gets \mathit{FS} \cup \{(Z,z_0,Z_\mathsf{OUT})\}$
            \label{l:addFragment}
            \State $R \gets R \cup Z$ \label{l:addToR1}
			\EndFor \label{l:forE}
			\State \Return {$\mathit{FS}$} \Comment{set of fragments}
		    \EndFunction
		\end{algorithmic}
	\end{small}
\end{algorithm}

\begin{algorithm}[t]
\caption{Traversal of pDTMC induced graph} \label{algorithm:fragmentgrow}
	\begin{small}
		\renewcommand{\baselinestretch}{1}
		\begin{algorithmic}[1]
			\Function{Traverse}{$\mathcal{G}(V,E), w, T, \mathit{FS}, R, Z, \mathit{isInput}$}
			\If{$\neg \mathit{isInput}$} \Comment{$w$ is not the fragment's input} \label{l:ifInputS}
    			\State $I \gets \{i \;|\; i\notin Z \wedge (i,w) \in E\}$ \label{l:inputs}
    			\If{$I \subseteq V\setminus R$} \label{l:wAddedIf} 
    			    \State $T.\textsc{Push}(I)$\label{l:wAdded}
    		    \Else \label{l:wFragment2S}
    		        \State $\mathit{FS} \gets \mathit{FS} \cup \{(\{w\},w,\{w\})\}$ \label{l:degenerate3}
    		        \State $R\gets R \cup \{w\}$ \label{l:addToR2}
    		        \State \Return \label{l:wFragment2E}
    			\EndIf
    		\EndIf  \label{l:ifInputE}

    			\State $O \gets \{o \;|\; o\notin Z \wedge (w,o) \in E\}$ \label{l:OutputS}
    			
    		\If{$O \not\subseteq R$} \label{l:notSubsetOfR}
    				\ForAll{$o \in O$} \label{l:OutputMid}
        			    \If{$o\notin R$} 
        			        \State $T.\textsc{Push}(o)$ \label{l:pushO}
        			     \Else
                            \State $\textsc{Restructure}(\mathcal{G},w)$ \Comment{Section~\ref{ssec:modelRestructuring}} \label{l:restructure2}
        			    \EndIf
        			\EndFor
        	   \EndIf 
        	    \label{l:OutputE}
		    \EndFunction

		\end{algorithmic}
	\end{small}
\end{algorithm}

\begin{figure*}
     \centering
     \begin{subfigure}[b]{0.85\textwidth}
         \centering
         \includegraphics[width=\textwidth]{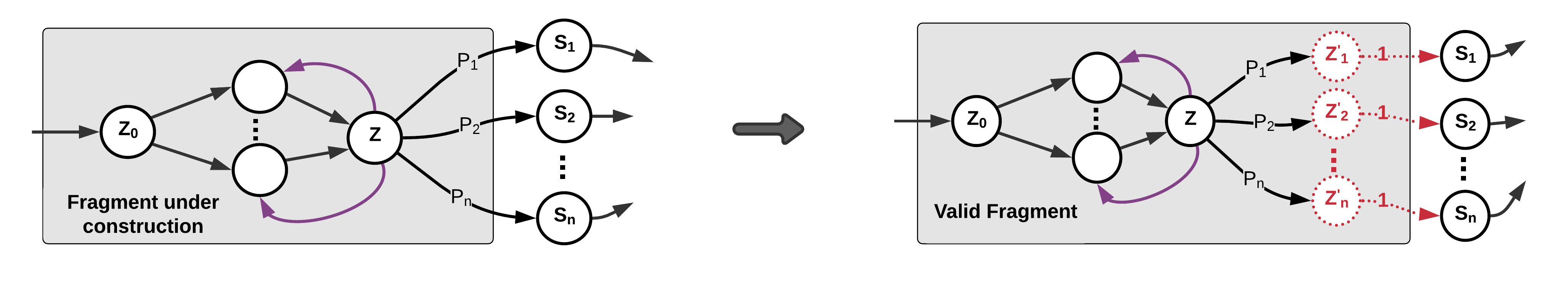}
         
         \vspace*{-3mm}
         \caption{Auxiliary state insertion to force formation of fragment containing the new states among its output states}
         \label{fig:trick1}
     \end{subfigure}
     \hfill
     
     \vspace*{1.5mm}
     \begin{subfigure}[b]{0.85\textwidth}
         \centering
         \includegraphics[width=\textwidth]{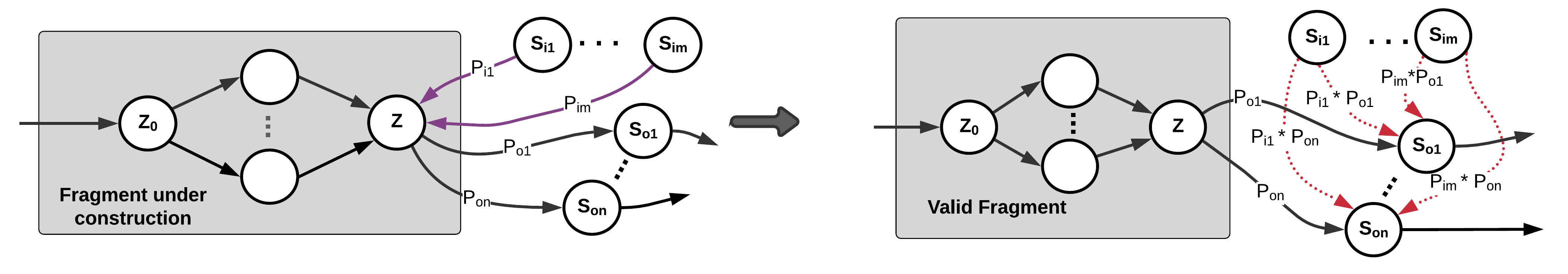}

         \vspace*{-3mm}
         \caption{Transition replacement to force creation of output fragment state}
         \label{fig:trick2}
     \end{subfigure}
        \caption{Model restructuring techniques supporting fragment formation}
        \label{fig:ModelReconstruction}
\end{figure*}

Due to space constraints, the full analysis of Algorithms~\ref{algorithm:fragmentation} and~\ref{algorithm:fragmentgrow} is provided as supplementary material on our project webpage. Here, we note that the algorithms are guaranteed to produce a set of valid fragments, comprising: (i)~the degenerate, single-state fragments created in lines~\ref{l:satisfy} and~\ref{l:degenerate2} of \textsc{Fragmentation} and in line~\ref{l:degenerate3} of \textsc{Traverse}; (ii)~the fragments that ``pass'' the validation from line~\ref{l:validfragmentS} of \textsc{Fragmentation}, which we assume correct. Furthermore, both algorithms terminate. \textsc{Traverse} terminates because each of its statements (including the assembly of the vertex sets $I$ and $O$, and its only for loop) operate with a finite number of vertices. \textsc{Fragmentation} terminates because: (i)~each iteration of its for loop adds at least one vertex (i.e., $z_0$) to $R$ in line~\ref{l:addToR1} until $V\setminus R = \{\}$ in line~\ref{l:forS} (since $V$ is a finite set) and the loop terminates; (ii)~its while loop terminates since it iterates over the elements of stack $T$ that can only contain one instance of vertices from the finite set $V$ and is therefore finite too; and (iii)~\textsc{Restructure} invocations can only add a finite number of vertices to $V$, as we show in the next section.

\setlength{\textfloatsep}{\textfloatsepsave}

\subsection{Model Restructuring to Aid Fragment Formation}\label{ssec:modelRestructuring}

While the model fragmentation from \S\ref{ssec:algorithm} is guaranteed to produce valid fragments, the success of \approach\ also depends on these fragments being neither too small nor too large. Small fragments may yield a large abstract model that may be unfeasible to analyse using standard PMC (Fig.~\ref{fig:approach}); however, in our experience, the pDTMCs whose PMC we want to speed up comprise many loops (cf.~Fig~\ref{fig:fxExample}) that preclude the formation of only small fragments. In contrast, fragments that are too large for standard PMC are more likely to be obtained. As such, \approach\ uses the threshold $\alpha$ to decide when to attempt to force the formation of a fragment in line~\ref{l:restructure} from Algorithm~\ref{algorithm:fragmentation}, and further attempts to enable fragment formation in line~\ref{l:restructure2} from Algorithm~\ref{algorithm:fragmentgrow}, in the two scenarios below. 

1) When a state $z$ of a fragment under construction has both outgoing transitions to inner fragment states and $n\geq 1$ outgoing transitions (of probabilities $p_1$, $p_2$, \ldots, $p_n$) to states $s_1$, $s_2$, \ldots $s_n$ not in the fragment (Fig~\ref{fig:trick1}, left). In this case, the addition of one auxiliary state $z'_i$ for each outgoing transition to a state $s_i$, $i=1,2,\ldots, n$ (Fig~\ref{fig:trick1}, right) enables the formation of a fragment that includes the auxiliary states as output states. Each auxiliary state $z'_i$ has an incoming transition of probability $p_i$ from state $z$, and an outgoing transition of probability $1$ going to state $s_i$.\footnote{Using a single auxiliary state $z'$ obtained by combining states $z'_1$, $z'_2$, \ldots, $z'_n$ is also possible, but leads to more complex expressions for the new transition probabilities, so we did not opt for it in the current version of \approach.}

2) When a non-input state $z$ of a fragment under construction has $m\geq 1$ incoming transitions (of probabilities $p_{i1}$, $p_{i2}$, \ldots, $p_{im}$) from states $s_{i1}$, $s_{i2}$, \ldots $s_{im}$ outside the fragment, and $n\geq 1$ outgoing transitions (of probabilities $p_{o1}$, $p_{o2}$, \ldots, $p_{on}$) to states $s_{o1}$, $s_{o2}$, \ldots $s_{on}$ not in the fragment (Fig~\ref{fig:trick2}, left). In this case, each of the $m$ transitions from a state $s_{ij}$ to $z$, $j=1,2,\ldots,m$, can be replaced by $n$ transitions from state $s_{ij}$ to states $s_{o1}$, $s_{o2}$, \ldots $s_{on}$ (Fig~\ref{fig:trick2}, right), where the  probability of transition from $s_{ij}$ to $s_{ok}$, $k=1,2,\ldots,n$, is set to $p_{ij}p_{ok}$.

Note that the restructuring process is invoked halfway through the construction of a fragment, when $Z_\mathsf{OUT}$ has not yet been finalised. Thus, the model restructuring process can apply to any state $z$ that satisfies one of the above scenarios and that we may want to include into $Z_\mathsf{OUT}$, but cannot because one of the constraints from the definition of a fragment is violated.

\begin{theorem} 
Applying the model restructuring techniques from Fig.~\ref{fig:ModelReconstruction} to a pDTMC does not affect its reachability properties.
\end{theorem}

\vspace*{-3mm}
\begin{proof} 
Consider the sets $\Pi$ and $\Pi'$ of all paths (i.e., sequences of possible states and state transitions) that satisfy a reachability property $P$ of a pDTMC $D=(S,s_0,\textbf{P}, L)$ before and after model restructuring is applied to change it to $D'=(S',s_0,\textbf{P}', L)$, respectively. 
According to the semantics of PCTL~\cite{bianco_alfaro_1995,hansson1994logic}, we need to show that $\mathrm{Pr}_{s_0}(\Pi)=\mathrm{Pr}'_{s_0}(\Pi')$, where $\mathrm{Pr}_{s_0}$ is a probability measure defined over all paths $\pi=s_0s_1s_2\ldots s_n$ starting in the initial state $s_0$ of pDTMC $D$ such that $\mathrm{Pr}_{s_0}(\pi)=\prod_{i=0}^{n-1}\mathbf{P}(s_i,s_{i+1})$, and $\mathrm{Pr}'_{s_0}$ is a similarly defined probability measure for $D'$. We focus on the paths that differ between $\Pi$ and $\Pi'$, and show that $\mathrm{Pr}_{s_0}(\Pi\setminus \Pi')=\mathrm{Pr}'_{s_0}(\Pi'\setminus \Pi)$ for each  technique in turn.
    
    1) A path from $\Pi\setminus \Pi'$ has the form $\pi=s_0\omega_1 zs_i\omega_2$ for some $i\in\{1,2,\ldots, n\}$ and subpaths $\omega_1$, $\omega_2$, with $\omega_2$ ending in one of the states from the reachability state set. For any such path $\pi$, there is a corresponding path $\pi'=s_0\omega_1 zz'_is_i\omega_2 \in \Pi'\setminus \Pi$, \emph{and the other way around}. We have $\mathrm{Pr}_{s_0}(\pi)=\mathrm{Pr}_{s_0}(s_0\omega_1)\mathbf{P}(z,s_i)\mathrm{Pr}_{s_0}(s_i\omega_2)=\mathrm{Pr}_{s_0}(s_0\omega_1) p_i\mathrm{Pr}_{s_0}(s_i\omega_2)=\mathrm{Pr}'_{s_0}(s_0\omega_1)$ $\mathbf{P}'(z,z'_i)\mathbf{P}'(z'_i,s_i)$ $\mathrm{Pr}'_{s_0}(s_i\omega_2)=\mathrm{Pr}'_{s_0}(\pi')$, so the theorem holds for the first restructuring technique. 
    
    
    2) A path from $\Pi\setminus \Pi'$ has the form $\pi=s_0\omega_1s_{oj}zs_{ik}\omega_2$, with $j\in\{1,2,\ldots,m\}$,  $k\in\{1,2,\ldots,n\}$, and subpath $\omega_2$ ending in one of the states from the reachability state set. Path $\pi$ has a corresponding path $\pi'=s_0\omega_1s_{oj}s_{ik}\omega_2\in \Pi'\setminus \Pi$, \emph{and the other way around}, and it is straightforward to show that $\mathrm{Pr}_{s_0}(\pi)=\mathrm{Pr}'_{s_0}(\pi')$ since $\mathbf{P}(s_{ij},z)\mathbf{P}(z,s_{ok})=\mathbf{P'}(s_{ij},s_{ok})$, which completes the proof.
\end{proof}

From the two model restructuring techniques, only the first increases the number of pDTMC states -- with as many auxiliary states as there are ougoing transitions from state $z$ to states outside the fragment under construction. Because (i)~the pDTMC has a finite number of states, each with a finite number of outgoing transitions; and (ii)~the auxiliary states do not require the application of the first technique (as they have a single outgoing transition), it follows that the maximum number of auxiliary states that \approach\ may create is also finite.

\subsection{\emph{\approach} Application to the Running Example}\label{ssec:fpmcapplication}

Applied to the pDTMC and reachability property for the FX system from our running example (which current parametric model checkers cannot analyse, cf.~\S\ref{sec:example}), our \approach\ tool (run for $\alpha=6$) generated the model fragmentation from Fig.~\ref{fig:runningExampleApplication}. The five single-state and eight multi-state fragments (13~fragments in total) comprise 49~states and 83~transitions compared to the 29~states and 58~transitions of the original pDTMC (Fig.~\ref{fig:fxExample}), with the additional states and transitions due to the model restructuring techniques from \S\ref{ssec:modelRestructuring}. The end-to-end computation of a closed-form analytical model for the FX success probability took \approach\ 4.43s, and the algebraic formulae of this model contain 1,456 arithmetic operations and took 0.002s to evaluate in Matlab (on a computer with the specification from Section~\ref{ssec:experimentalSetup}). As for all \approach\ experiments presented in the paper, we carefully checked the correctness of our PMC formulae by ensuring that their evaluation for randomly generated combinations of parameter values produced the same numerical results (subject to negligible rounding errors) as those obtained by running the probabilistic model checkers PRISM~\cite{prism} and Storm~\cite{storm} to analyse the non-parametric Markov chains obtained by replacing the pDTMC parameters with these random values. 

\begin{figure}
	\centering
    \includegraphics[width=0.95\linewidth]{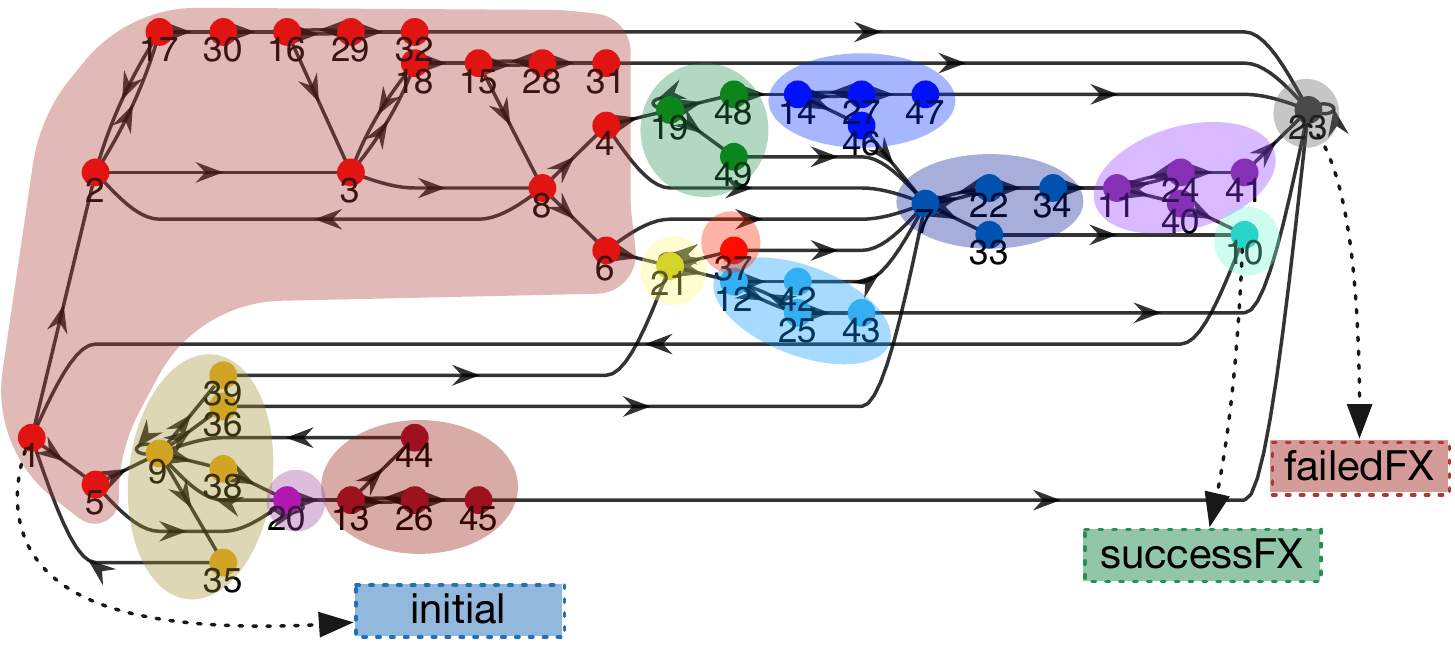}	
    \caption{Fragmentation of the pDTMC model from the running example (with the 13 fragments depicted in different colours)}
    \label{fig:runningExampleApplication}
    
    \vspace*{-2mm}
\end{figure}

\section{Implementation}\label{sec:implementation}

To ease the evaluation and adoption of \approach, we developed a prototype Java tool that implements Algorithms~\ref{algorithm:fragmentation} and~\ref{algorithm:fragmentgrow} from \S\ref{sec:approach}. The tool invokes the  model checkers PRISM~\cite{prism} and Storm~\cite{storm} to obtain the pDTMC transition probability matrix, and to compute the PMC formulae for the pDTMC fragments and abstract model, respectively. The open-source \approach\ prototype, the full experimental results summarised next, additional information about \approach\ and the case studies used for its evaluation are available on our project webpage at~\url{https://github.com/xinwei2124/fPMC_ICSE}. 

\section{Evaluation}\label{sec:evaluation}

\subsection{Research Questions}\label{ssec:rqs}

We carried out experiments to answer the following three research questions (RQs). 

\vspace{0.15cm}
\noindent
\textbf{RQ1 (Effectiveness): Can model fragmentation improve PMC efficiency and extend its applicability?} 
We assessed whether \approach\ can speed up PMC compared to PRISM~\cite{prism} and Storm~\cite{storm}, and whether our approach enables the analysis of models that these model checkers cannot handle.

\vspace{0.15cm}
\noindent
\textbf{RQ2 (Scalability): 
How does the number of parameters in a pDTMC model affect the ability of \approach\ to fragment the model and compute closed-form formulae?} As the number of pDTMC model parameters is a factor influencing the performance of current PMC solutions, we 
investigated how \approach\ performs as the number of such parameters increases.

\vspace*{0.15cm}
\noindent
\textbf{RQ3 (Configurability): 
What is the impact of the hyperparameter $\alpha$ on \approach?}
We examined how different threshold $\alpha$ values, i.e., the only \approach\ hyperparameter (cf. Algorithm~\ref{algorithm:fragmentation}), affect
 \approach\ in terms of the number of operations in the computed closed-form analytical model and execution time.

\subsection{Experimental Setup}\label{ssec:experimentalSetup}

We evaluated \approach\ for 
three software systems and processes taken from related research~\cite{RaduePMC,ghezzi2013model,hajnal2019data,classen2010model,Gerasimou2015:ASE} and belonging to different application domains. 
We selected these systems because 
(i) their Markov models include hyperparameters that can be tuned to devise pDTMCs with various numbers of states $S$ and transition probability matrices $\textbf{P}$; and
(ii) without changing the pDTMC structure, several non-trivial transition probabilities in $\textbf{P}$ can become parameters, thus increasing the number of parameters within the model. 
Due to space constraints, we only provide brief descriptions of these systems; 
full details are available from~\cite{RaduePMC,ghezzi2013model,hajnal2019data,classen2010model,Gerasimou2015:ASE}. 

\vspace*{1.5mm}
\noindent\textbf{FX System.} 
We have presented the high-level overview of the FX system in Section~\ref{sec:example} and the pDTMC corresponding to the sequential execution strategy with retry (SEQ\_R) in Fig.~\ref{fig:fxmodel}.
We also consider the strategies below for executing between 1--5 functionally-equivalent service implementations per operation:

\squishlistBigLabel
    \item [SEQ:]
    The services are invoked in order, stopping after the first successful invocation or after the last service fails.
    
    \item [PAR:]
    All services are invoked in parallel (i.e., simultaneously), and the operation uses the result returned by the first service terminating successfully.
    
    \item [PROB:] 
    A probabilistic selection is made among the available services with the total probability being equal to 1.

    \item[PROB\_R:]
    Similar to PROB, but if the selected service fails, it is retried with a given probability (as in SEQ\_R).
\squishend

\vspace*{1mm}
\noindent\textbf{PL System.} 
This product line system (PL)~\cite{ghezzi2013model,classen2010model} models the process in various vending machines that enable a user to insert coins and select a beverage, based on which the vending machine delivers the beverage and, if needed, gives changes. 
The features of this system comprises the beverage type (soda, tea, or both), payment mode (cash or free) and taste preference  (e.g., add lemon, sugar) enabling the derivation of vending machines, and accordingly the definition of pDTMCs, whose structures contain between 4 and 22 features. 

\vspace*{1.5mm}
\noindent\textbf{COM Process.} 
We consider a communication (COM) process~\cite{hajnal2019data}  among $n$ identical individuals, inspired by the way that honeybees emit an alarm pheromone to recruit workers and protect their colonies from intruders.  
Due to the self-destructive defence behaviour in social insects, the recruited workers die after completing their defence actions.
Hence, 
a balance between efficient defence and preservation of a critical workers mass can be found. The induced pDTMC is a stochastic population model with $n$ parameters. The quantitative analysis of such stochastic models of multi-agent systems is often challenging because the dependencies among the agents within the population make the models complex. 

\vspace{1.5mm}
We compare \approach\ against the leading PMC model checkers PRISM (version 4.6) and Storm (version 1.5.1), both with their default settings. 
For a fair comparison, we ensured that both PRISM and Storm can process at least the simpler pDTMCs of these systems. 
We run each experiment using \approach, PRISM and Storm, and observe
(i) the time required to compute the PMC formulae, with a 60-minute timeout (the lower the execution time the better); and 
(ii) the number of arithmetic operations in the devised formulae (formulae with fewer arithmetic operations can be evaluated faster). 

All experiments were performed on a MacBook Pro (early 2015) with 2.7GHz dual Core Intel i5 processor and 8GB RAM. 
The source code, the full Markov models and reachability properties, and all experimental results are publicly available at \url{https://github.com/xinwei2124/fPMC_ICSE}. 

\subsection{Results \& Discussion}\label{ssec:results}


\begin{table*}[t!]
\setlength{\tabcolsep}{4pt}
\renewcommand{\arraystretch}{0.95}
\caption{
Parametric model checking for the FX system showing the execution time, the number of arithmetic operations of the closed-form formulae and the number of fragments devised by \approach.} 
\centering
\begin{tabular}{cc|cc|cccc|cccc|cc}
\hline
\multirow{2}{*}{\textbf{STG}} &
\multirow{2}{*}{\textbf{\#SRV}} &
  \multirow{2}{*}{\textbf{\#S}} &
  \multirow{2}{*}{\textbf{\#T}} &
  \multicolumn{4}{c|}{\textbf{Time (s)}} &
  \multicolumn{4}{c|}{\textbf{\#arithmetic operations}}&
  \multicolumn{2}{c}{\textbf{\#fragments}}
  \\
\multicolumn{2}{c|}{} &
  & &
  \textbf{\approach} &
  \textbf{\approach\ ($\alpha\!\!=\!\!\infty$}) &
  \textbf{Storm} &
  \multicolumn{1}{c|}{\textbf{PRISM}} &
  \textbf{\approach} &
  \textbf{\approach\ ($\alpha\!\!=\!\!\infty$}) &
  \textbf{Storm} &
  \textbf{PRISM} &
  \textbf{\approach} &
  \textbf{\approach\ ($\alpha\!=\!\infty$})
  \\ \hline
\multirow{5}{*}{\textbf{SEQ}}    & 1 & 11&22   & 2.588    & 2.814   & 0.017   & \multicolumn{1}{c|}{0.459} & 228    & 228    & 157    & 160   &  4 & 4\\ 

                                 & 2 & 17&34   & 2.684    & 3.448   & 2.31    & \multicolumn{1}{c|}{51.46} & 479    & 479    & 35488  & 34074  & 6 & 6  \\
                                 & 3 & 23&46   & 3.563    & 5.212   & T       & \multicolumn{1}{c|}{T}     & 1374   & 1374   & --    & --  & 7 & 7 \\
                                 & 4 & 29&58   & 4.366    & 5.195   & T       & \multicolumn{1}{c|}{T}     & 3304   & 3304   & --    & --   & 9 & 9\\
                                 & 5 & 35&70   & 7.973    & 7.770   & T       & \multicolumn{1}{c|}{T}     & 6481   & 6481   & --    & --   & 10 & 10\\ \hline
\multirow{4}{*}{\textbf{PAR}}    & 2 & 40&36   & 3.717    & 3.834   & 1.743   & \multicolumn{1}{c|}{1.321} & 909    & 909    & 35519  & 32134 & 18 & 18\\
                                 & 3 & 64&111  & 5.369    & 6.751   & T       & \multicolumn{1}{c|}{T}     & 7952   & 7952   & --    & -- & 26 & 26  \\
                                 & 4 & 112&207 & 14.274   & 101.949 & T       & \multicolumn{1}{c|}{T}     & 6062   & 68574  & --    & --   & 27 & 41\\
                                 & 5 & 208&399 & 171.903  & T       & T       & \multicolumn{1}{c|}{T}     & 24502  & --    & --    & --   & 43 & --\\ \hline
\multirow{4}{*}{\textbf{PROB}}   & 2 & 23&46   & 3.385    & 4.303   & 0.377   & \multicolumn{1}{c|}{8.352} & 743    & 743    & 6517   & 20991 & 8 & 8 \\
                                 & 3 & 29&64   & 4.533    & 5.618   & 3.735   & \multicolumn{1}{c|}{T}     & 1970   & 1970   & 47812  & -- & 10 & 10   \\
                                 & 4 & 35&82   & 6.253    & 7.322   & 34.836  & \multicolumn{1}{c|}{T}     & 4136   & 4136   & 196985 & --  & 12 & 12   \\
                                 & 5 & 41&100  & 10.694   & 13.016  & 619.505 & \multicolumn{1}{c|}{T}     & 7508   & 7508   & 593426 & --   & 14 & 14  \\ \hline
\multirow{4}{*}{\textbf{SEQ\_R}}  & 2 & 29&58   & 5.340    & 216.792 & T       & \multicolumn{1}{c|}{T}     & 3068   & 124854 & --    & --  & 9 & 3 \\
                                 & 3 & 41&82   & 232.811  & T       & T       & \multicolumn{1}{c|}{T}     & 37317  & --    & --    & --   & 12 & --  \\
                                 & 4 & 53&106  & 173.624  & T       & T       & \multicolumn{1}{c|}{T}     & 131336 & --    & --    & --   & 30 & --  \\
                                 & 5 & 65&130  & 1561.751 & T       & T       & \multicolumn{1}{c|}{T}     & 385893 & --    & --    & --   & 36 & --  \\ \hline
\multirow{4}{*}{\textbf{PROB\_R}} & 2 & 29&58   & 4.790    & 34.037  & 54.822  & \multicolumn{1}{c|}{T}     & 1735   & 27251  & 84449  & --  & 15 & 3   \\
                                 & 3 & 35&75   & 15.261   & T       & T       & \multicolumn{1}{c|}{T}     & 4832   & --    & --    & --   & 19 & --  \\
                                 & 4 & 41&93   & 57.187   & T       & T       & \multicolumn{1}{c|}{T}     & 10437  & --    & --    & --   & 23 & --  \\
                                 & 5 & 47&111  & 179.623  & T       & T       & \multicolumn{1}{c|}{T}     & 19306  & --    & --    & --  & 27 & --   \\ 
                                 \hline
\multicolumn{14}{l}{\textbf{Notation} -- STG: strategy adopted for the functionally-equivalent service implementations per FX operation; \#SRV: available service implementations }\\

\multicolumn{14}{l}{per operation; \#S: pDTMC states; \#T: pDTMC probabilistic transitions; 
\approach: fPMC with $\alpha\!=\!15$ except from SEQ\_R \#SRV=4,5 where $\alpha\!=\!3$;}\\

\multicolumn{14}{l}{\approach $\;(\alpha\!=\!\infty)$:  \approach\ variant where the invocation of the $\textsc{Restructure}$ function in Algorithm~\ref{algorithm:fragmentation} is disabled; T: timeout--no result returned in 60min.}\\
\end{tabular}
\label{Table:FXmodel}
\end{table*}

\begin{table*}[h!]
\setlength\tabcolsep{5pt}
\renewcommand{\arraystretch}{0.95}
\caption{
Parametric model checking for four variants of the PL system (with 4,16,18 and 22 features) showing the execution time and the number of arithmetic operations (\#OP) of the computed closed-form formulae for different percentages of pDTMC model parameters. 
For the pDTMC with 4,16 and $\!$ 18 features $\alpha\!\!=\!10$, and for the pDTMC with 22 features $\alpha\!\!=\!5$.
}
\footnotesize
\centering
\begin{tabular}{crlccccccccccc}
\hline
 &
  \multicolumn{2}{c}{} &
  \multicolumn{11}{c}{\textbf{Percentage of parameters in the pDTMC}} \\ \cline{4-14} 
\multirow{-2}{*}{\textbf{pDTMC}} &
  \multicolumn{2}{c}{\multirow{-2}{*}{\textbf{Model checker}}} &
  \textbf{P01} &
  \textbf{P10} &
  \textbf{P20} &
  \textbf{P30} &
  \textbf{P40} &
  \textbf{P50} &
  \textbf{P60} &
  \textbf{P70} &
  \textbf{P80} &
  \textbf{P90} &
  \textbf{P100} \\ \hline
 &
   &
  \textbf{Time (s)} &
  \cellcolor[HTML]{ECECEC}40.382 &
  \cellcolor[HTML]{ECECEC}47.838 &
  \cellcolor[HTML]{ECECEC}57.045 &
  \cellcolor[HTML]{ECECEC}58.278 &
  \cellcolor[HTML]{ECECEC}57.101 &
  \cellcolor[HTML]{ECECEC}69.933 &
  \cellcolor[HTML]{ECECEC}79.454 &
  \cellcolor[HTML]{ECECEC}89.109 &
  \cellcolor[HTML]{ECECEC}91.013 &
  \cellcolor[HTML]{ECECEC}93.623 &
  \cellcolor[HTML]{ECECEC}93.063 \\
 &
  \multirow{-2}{*}{\textbf{\approach}} &
  \textbf{\#OP} &
  38681 &
  46975 &
  50794 &
  54903 &
  55456 &
  67649 &
  75120 &
  94445 &
  94489 &
  94757 &
  90277 \\ \cline{2-14} 
 &
   &
  \textbf{Time (s)} &
  \cellcolor[HTML]{ECECEC}0.002 &
  \cellcolor[HTML]{ECECEC}0.01 &
  \cellcolor[HTML]{ECECEC}0.221 &
  \cellcolor[HTML]{ECECEC}3.195 &
  \cellcolor[HTML]{ECECEC}32.000 &
  \cellcolor[HTML]{ECECEC}1674.020 &
  \cellcolor[HTML]{ECECEC}T &
  \cellcolor[HTML]{ECECEC}T &
  \cellcolor[HTML]{ECECEC}T &
  \cellcolor[HTML]{ECECEC}T &
  \cellcolor[HTML]{ECECEC}T \\
 &
  \multirow{-2}{*}{\textbf{Storm}} &
  \textbf{\#OP} &
  3 &
  772 &
  6560 &
  24048 &
  39805 &
  286349 &
  800693 &
  -- &
  -- &
  -- &
  -- \\ \cline{2-14} 
 &
   &
  \textbf{Time (s)} &
  \cellcolor[HTML]{ECECEC}0.215 &
  \cellcolor[HTML]{ECECEC}0.249 &
  \cellcolor[HTML]{ECECEC}0.321 &
  \cellcolor[HTML]{ECECEC}1.476 &
  \cellcolor[HTML]{ECECEC}5.088 &
  \cellcolor[HTML]{ECECEC}105.520 &
  \cellcolor[HTML]{ECECEC}T &
  \cellcolor[HTML]{ECECEC}T &
  \cellcolor[HTML]{ECECEC}T &
  \cellcolor[HTML]{ECECEC}T &
  \cellcolor[HTML]{ECECEC}T \\
\multirow{-6}{*}{\textbf{\begin{tabular}[c]{@{}c@{}}4 \\ Features \\ \#S=92 \\ \#T=167\\\#params=75\end{tabular}}} &
  \multirow{-2}{*}{\textbf{PRISM}} &
  \textbf{\#OP} &
  1 &
  29 &
  578 &
  5738 &
  42062 &
  475809 &
  -- &
  -- &
  -- &
  -- &
  -- \\ \hline
 &
   &
  \textbf{Time (s)} &
  \cellcolor[HTML]{ECECEC}18.335 &
  \cellcolor[HTML]{ECECEC}48.179 &
  \cellcolor[HTML]{ECECEC}52.260 &
  \cellcolor[HTML]{ECECEC}20.162 &
  \cellcolor[HTML]{ECECEC}20.077 &
  \cellcolor[HTML]{ECECEC}23.309 &
  \cellcolor[HTML]{ECECEC}22.547 &
  \cellcolor[HTML]{ECECEC}22.783 &
  \cellcolor[HTML]{ECECEC}23.634 &
  \cellcolor[HTML]{ECECEC}23.778 &
  \cellcolor[HTML]{ECECEC}23.790 \\
 &
  \multirow{-2}{*}{\textbf{\approach}} &
  \textbf{\#OP} &
  55482 &
  79415 &
  81293 &
  86648 &
  86684 &
  107475 &
  107166 &
  107339 &
  107734 &
  107751 &
  107769 \\ \cline{2-14} 
 &
   &
  \textbf{Time (s)} &
  \cellcolor[HTML]{ECECEC}0.001 &
  \cellcolor[HTML]{ECECEC}0.361 &
  \cellcolor[HTML]{ECECEC}15.709 &
  \cellcolor[HTML]{ECECEC}23.489 &
  \cellcolor[HTML]{ECECEC}31.624 &
  \cellcolor[HTML]{ECECEC}164.624 &
  \cellcolor[HTML]{ECECEC}T &
  \cellcolor[HTML]{ECECEC}T&
  \cellcolor[HTML]{ECECEC}T &
  \cellcolor[HTML]{ECECEC}T &
  \cellcolor[HTML]{ECECEC}T \\
 &
  \multirow{-2}{*}{\textbf{Storm}} &
  \textbf{\#OP} &
  3 &
  772 &
  6560 &
  24048 &
  39805 &
  286349 &
  -- &
  -- &
  -- &
  -- &
  -- \\ \cline{2-14} 
 &
   &
  \textbf{Time (s)} &
  \cellcolor[HTML]{ECECEC}0.353 &
  \cellcolor[HTML]{ECECEC}14.917 &
  \cellcolor[HTML]{ECECEC}T &
  \cellcolor[HTML]{ECECEC}T &
  \cellcolor[HTML]{ECECEC}T &
  \cellcolor[HTML]{ECECEC}T &
  \cellcolor[HTML]{ECECEC}T &
  \cellcolor[HTML]{ECECEC}T &
  \cellcolor[HTML]{ECECEC}T &
  \cellcolor[HTML]{ECECEC}T &
  \cellcolor[HTML]{ECECEC}T \\
\multirow{-6}{*}{\textbf{\begin{tabular}[c]{@{}c@{}}16\\ Features \\ \#S=110 \\ \#T=193\\\#params=83\end{tabular}}} &
  \multirow{-2}{*}{\textbf{PRISM}} &
  \textbf{\#OP} &
  1 &
  606 &
  -- &
  -- &
  -- &
  -- &
  -- &
  -- &
  -- &
  -- &
  -- \\ \hline
 &
   &
  \textbf{Time (s)} &
  \cellcolor[HTML]{ECECEC}13.677 &
  \cellcolor[HTML]{ECECEC}13.828 &
  \cellcolor[HTML]{ECECEC}13.719 &
  \cellcolor[HTML]{ECECEC}15.286 &
  \cellcolor[HTML]{ECECEC}14.294 &
  \cellcolor[HTML]{ECECEC}14.431 &
  \cellcolor[HTML]{ECECEC}14.368 &
  \cellcolor[HTML]{ECECEC}14.844 &
  \cellcolor[HTML]{ECECEC}17.464 &
  \cellcolor[HTML]{ECECEC}17.347 &
  \cellcolor[HTML]{ECECEC}17.471 \\
 &
  \multirow{-2}{*}{\textbf{\approach}} &
  \textbf{\#OP} &
  25703 &
  27643 &
  27643 &
  31499 &
  31672 &
  35523 &
  35528 &
  39400 &
  70816 &
  76303 &
  76388 \\ \cline{2-14} 
 &
   &
  \textbf{Time (s)} &
  \cellcolor[HTML]{ECECEC}0.02 &
  \cellcolor[HTML]{ECECEC}0.043 &
  \cellcolor[HTML]{ECECEC}0.116 &
  \cellcolor[HTML]{ECECEC}0.554 &
  \cellcolor[HTML]{ECECEC}18.773 &
  \cellcolor[HTML]{ECECEC}911.039 &
  \cellcolor[HTML]{ECECEC}T &
  \cellcolor[HTML]{ECECEC}T &
  \cellcolor[HTML]{ECECEC}T &
  \cellcolor[HTML]{ECECEC}T &
  \cellcolor[HTML]{ECECEC}T \\
 &
  \multirow{-2}{*}{\textbf{Storm}} &
  \textbf{\#OP} &
  4 &
  104 &
  206 &
  13101 &
  75459 &
  1179040 &
  -- &
  -- &
  -- &
  -- &
  -- \\ \cline{2-14} 
 &
   &
  \textbf{Time (s)} &
  \cellcolor[HTML]{ECECEC}0.343 &
  \cellcolor[HTML]{ECECEC}0.963 &
  \cellcolor[HTML]{ECECEC}1.084 &
  \cellcolor[HTML]{ECECEC}2.499 &
  \cellcolor[HTML]{ECECEC}41.665 &
  \cellcolor[HTML]{ECECEC}86.148 &
  \cellcolor[HTML]{ECECEC}188.956 &
  \cellcolor[HTML]{ECECEC}T &
  \cellcolor[HTML]{ECECEC}T &
  \cellcolor[HTML]{ECECEC}T &
  \cellcolor[HTML]{ECECEC}T \\
\multirow{-6}{*}{\textbf{\begin{tabular}[c]{@{}c@{}}18 \\ Features \\ \#S=104 \\ \#T=183\\\#params=79\end{tabular}}} &
  \multirow{-2}{*}{\textbf{PRISM}} &
  \textbf{\#OP} &
  4 &
  82 &
  177 &
  10458 &
  82654 &
  1218965 &
  3909823 &
  -- &
  -- &
  -- &
  -- \\ \hline
 &
   &
  \textbf{Time (s)} &
  \cellcolor[HTML]{ECECEC}24.268 &
  \cellcolor[HTML]{ECECEC}37.804 &
  \cellcolor[HTML]{ECECEC}46.97 &
  \cellcolor[HTML]{ECECEC}45.16 &
  \cellcolor[HTML]{ECECEC}44.217 &
  \cellcolor[HTML]{ECECEC}42.943 &
  \cellcolor[HTML]{ECECEC}45.598 &
  \cellcolor[HTML]{ECECEC}44.338 &
  \cellcolor[HTML]{ECECEC}45.306 &
  \cellcolor[HTML]{ECECEC}47.015 &
  \cellcolor[HTML]{ECECEC}46.467 \\
 &
  \multirow{-2}{*}{\textbf{\approach}} &
  \textbf{\#OP} &
  116302 &
  187109 &
  210387 &
  210456 &
  210464 &
  210464 &
  210473 &
  222304 &
  222304 &
  260538 &
  260465 \\ \cline{2-14} 
 &
   &
  \textbf{Time (s)} &
  \cellcolor[HTML]{ECECEC}0.002 &
  \cellcolor[HTML]{ECECEC}0.035 &
  \cellcolor[HTML]{ECECEC}3.53 &
  \cellcolor[HTML]{ECECEC}42.334 &
  \cellcolor[HTML]{ECECEC}793.458 &
  \cellcolor[HTML]{ECECEC}T &
  \cellcolor[HTML]{ECECEC}T &
  \cellcolor[HTML]{ECECEC}T &
  \cellcolor[HTML]{ECECEC}T &
  \cellcolor[HTML]{ECECEC}T &
  \cellcolor[HTML]{ECECEC}T \\
 &
  \multirow{-2}{*}{\textbf{Storm}} &
  \textbf{\#OP} &
  4 &
  104 &
  206 &
  13101 &
  75459 &
  -- &
  -- &
  -- &
  -- &
  -- &
  -- \\ \cline{2-14} 
 &
   &
  \textbf{Time (s)} &
  \cellcolor[HTML]{ECECEC}0.548 &
  \cellcolor[HTML]{ECECEC}1.821 &
  \cellcolor[HTML]{ECECEC}53.239 &
  \cellcolor[HTML]{ECECEC}T &
  \cellcolor[HTML]{ECECEC}T &
  \cellcolor[HTML]{ECECEC}T &
  \cellcolor[HTML]{ECECEC}T &
  \cellcolor[HTML]{ECECEC}T &
  \cellcolor[HTML]{ECECEC}T &
  \cellcolor[HTML]{ECECEC}T &
  \cellcolor[HTML]{ECECEC}T \\
\multirow{-6}{*}{\textbf{\begin{tabular}[c]{@{}c@{}}22 \\ Features \\ \#S=115 \\ \#T=198\\\#params=83\end{tabular}}} &
  \multirow{-2}{*}{\textbf{PRISM}} &
  \textbf{\#OP} &
  4 &
  59 &
  1225 &
  -- &
  -- &
  -- &
  -- &
  -- &
  -- &
  -- &
  -- \\ \hline
     \multicolumn{14}{l}{\textbf{Notation} -- \#S: pDTMC states; \#T: pDTMC probabilistic transitions; \#params: total number of pDTMC parameters; $\!P01,\!...,\!P100$: percentage of pDTMC}\\
  \multicolumn{14}{l}{parameters maintained -- the others are set to random values $\in [0,1]$ preserving the pDTMC structure; T: timeout--no result returned within 60 minutes.}
\end{tabular}
\label{Table:PLmodel}

\vspace*{-1.5mm}
\end{table*}

\noindent
\textbf{RQ1 (Effectiveness).} 
Table~\ref{Table:FXmodel} shows  the execution time, the number of arithmetic operations of the computed closed-form formulae and the number of fragments devised by fPMC using FX system variants with different execution strategies and number of functionally-equivalent service implementations per operation. 
The size of the derived pDTMCs ranges from~11 states and 22 transitions (SEQ with 1 service implementation) to 208 states and 399 transitions (PAR with 5 services). 

For all pDTMCs of FX variants, \approach\ succeeded in computing all closed-form formulae within the 60 minutes timeframe (\approach\ column) taking $2.5s$ for the simplest models,  and just under 240s for most models with the exception of SEQ\_R with 5 services for which \approach\ took $\approx\!1560s$.
In contrast, Storm computed the formulae for eight of 21 pDTMCs ($\approx\!38\%$) with the majority being the simplest models across all execution strategies, except from PROB variants for which Storm produced all formulae.  
Finally, PRISM computed the PMC formulae for four of the 21 pDTMCs ($\approx\!20\%$), which again were the pDTMCs with the fewest states and transitions.

\begin{table}[t]
\setlength\tabcolsep{4.0pt}
\renewcommand{\arraystretch}{0.9}
\caption{
Parametric model checking of reachability properties $P1,\! ...,\!P21$ from~\cite{hajnal2019data} for the COM process with $n\!=\!20$ individuals (parameters) and $\alpha\!\!=\!\!50$, showing the execution time and the number of arithmetic operations of the derived formulae.
The pDTMC has 234 states and 444 transitions.
}
\begin{tabular}{cccccccc}
\hline
\multirow{2}{*}{\textbf{P\#}} &
  \multicolumn{3}{c}{\textbf{Time (s)}} &
  \multicolumn{4}{c}{\textbf{$\!\!\!\!$\#arithmetic operations (M:$\!$ Megabytes)}} \\ \cline{2-8} 
 &
  \textbf{\approach} &
  \textbf{Storm} &
  \textbf{PRISM} &
  \textbf{\approach} &
  \textbf{\begin{tabular}[c]{@{}c@{}}fPMC \\ (abstract\\model)\end{tabular}} & \textbf{Storm} & \textbf{PRISM} \\ \hline
\multicolumn{1}{c|}{1}  & 8.894 & 0.000   & \multicolumn{1}{c|}{0.747} & 98568   & 1      & 2             & 20      \\
\multicolumn{1}{c|}{2}  & 8.848 & 0.001   & \multicolumn{1}{c|}{0.648} & 98567   & 15     & 6             & 780     \\
\multicolumn{1}{c|}{3}  & 9.295 & 0.004   & \multicolumn{1}{c|}{0.847} & 98791   & 224    & 1352          & 11724   \\
\multicolumn{1}{c|}{4}  & 9.239 & 0.030   & \multicolumn{1}{c|}{1.241} & 100867  & 1561   & 16837         & 97814   \\
\multicolumn{1}{c|}{5}  & 9.146 & 0.427   & \multicolumn{1}{c|}{3.002} & 105314  & 6747   & 122879        & 544375  \\
\multicolumn{1}{c|}{6}  & 9.025 & 0.461   & \multicolumn{1}{c|}{4.867} & 118846  & 20279  & 610454        & 2M     \\
\multicolumn{1}{c|}{7}  & 8.340 & 1.768   & \multicolumn{1}{c|}{OM}    & 143612  & 45045  & 2M            & --     \\
\multicolumn{1}{c|}{8}  & 9.353 & 15.065  & \multicolumn{1}{c|}{OM}    & 175280  & 76713  & 5M            & --     \\
\multicolumn{1}{c|}{9}  & 8.898 & 18.870  & \multicolumn{1}{c|}{OM}    & 201098  & 102531 & 12M           & --     \\
\multicolumn{1}{c|}{10} & 8.802 & 94.334  & \multicolumn{1}{c|}{OM}    & 207676  & 109109 & 101M         & --     \\
\multicolumn{1}{c|}{11} & 9.162 & 73.880  & \multicolumn{1}{c|}{OM}    & 191660  & 93093  & 150M         & --     \\
\multicolumn{1}{c|}{12} & 8.976 & 316.619 & \multicolumn{1}{c|}{OM}    & 162203  & 63636  & --           & --     \\
\multicolumn{1}{c|}{13} & 9.160 & 130.994 & \multicolumn{1}{c|}{OM}    & 133069  & 34502  & 167M          & --     \\
\multicolumn{1}{c|}{14} & 8.876 & 39.222  & \multicolumn{1}{c|}{OM}    & 113092  & 14525  & 130M          & --     \\
\multicolumn{1}{c|}{15} & 9.247 & 22.488  & \multicolumn{1}{c|}{OM}    & 103173  & 4606   & 82M           & --     \\
\multicolumn{1}{c|}{16} & 9.223 & 8.401   & \multicolumn{1}{c|}{9.490} & 99646   & 1079   & 9M            & --     \\
\multicolumn{1}{c|}{17} & 9.077 & 3.853   & \multicolumn{1}{c|}{3.936} & 98763   & 196    & 3M            & 4M \\
\multicolumn{1}{c|}{18} & 8.911 & 0.587   & \multicolumn{1}{c|}{1.952} & 98595   & 28     & 1M            & 1M \\
\multicolumn{1}{c|}{19} & 9.353 & 0.128   & \multicolumn{1}{c|}{0.928} & 98568   & 1      & 263799        & 300712  \\
\multicolumn{1}{c|}{20} & 8.653 & 0.019   & \multicolumn{1}{c|}{0.638} & 98568   & 1      & 38832         & 41439   \\
\multicolumn{1}{c|}{21} & 9.124 & 0.004   & \multicolumn{1}{c|}{0.411} & 98568   & 1      & 2861          & 2870    \\ \hline
                        &       &         &                            &         &        &               &        
\vspace*{-3mm}
\end{tabular}
\label{Table:biologyModel}
\end{table}

Similar results were obtained for the four analysed PL system variants (with 4, 16, 18 and 22 features) whose results are shown in Table~\ref{Table:PLmodel}. 
Considering column P100, for instance, the most difficult case in which the transition probability matrix $\textbf{P}$ comprises only parameters (i.e., no transition is set to a constant value), \approach\ returned the closed-form formulae in less than $\approx\!\!94s$ in the worst case. Neither Storm nor PRISM succeeded in any of the four PL system variants for $P100$. 

While \approach\ outperforms both Storm and PRISM when handling complex models with several parameters, we noticed an interesting behaviour with simpler models (for FX and PL) and simpler properties (for COM).
In particular, for the simplest FX models with SEQ, PAR and PROB strategies and 1--2 services (cf. Table~\ref{Table:FXmodel}) mostly Storm, and occasionally PRISM, computed the formulae faster than \approach. This behaviour occurs for PL system models up to $P40$ in Table~\ref{Table:PLmodel} ($40\%$ of the total pDTMC parameters were maintained) and properties $P1\!\!-\!\!P7,P16\!\!-\!\!P21$  for the COM process (Table~\ref{Table:biologyModel}). 
Since both Storm and PRISM represent internally the induced pDTMCs with advanced data structures (e.g., sparse matrices, binary decision diagrams) and use sophisticated reachability analysis algorithms, this behaviour is logical for models and reachability properties that can exploit these features. 
As we have demonstrated, however, these features alone cannot handle reachability properties for pDTMCs with characteristics similar to those of the most complex models in Tables~\ref{Table:FXmodel}--\ref{Table:biologyModel}. 
This observation indicates the potential of a hybrid probabilistic model checker that uses \approach\ and Storm/PRISM interchangeably based on the pDTMC structure and reachability property.

Considering the complexity of the derived   closed-form formulae as a factor of the number  of  involved  arithmetic operations, we  observe that for all systems and all variants, the complexity increases as the model complexity (states and transitions)  increases.
For instance, for the FX system with SEQ\_R strategy (the motivating example) the arithmetic operations for the  \approach-computed formulae increase from 3068 to $\approx\!386\!K$ as the number of functionally-equivalent services increases from two to five.
In general, the \approach-produced formulae contain fewer operations than those produced by Storm or PRISM. 
This observation is evident in Table~\ref{Table:biologyModel}, where the closed-form analytical models derived from \approach\ include $98K \!\!-\!\!208K$ operations for all properties, whereas, in many cases, the formulae produced by Storm and PRISM are very large (between several megabytes and 167MB) and could not be analysed using our experimental machine.

These findings indicate that \approach, underpinned by the model fragmentation  algorithm, significantly speeds up the parametric model checking of reachability properties by several orders of magnitude, produces closed-form analytical models of modest complexity, and enables the analysis of pDTMCs that leading probabilistic model checkers cannot handle.
Also, the use of systems from different application domains, whose induced pDTMCs have distinct characteristics in terms of model structure and complexity, provides evidence supporting the general applicability of \approach. 
This conclusion is reinforced by the fact that the fragmentation is completely automated and does not require any domain knowledge or human intervention (cf. Algorithms~\ref{algorithm:fragmentation} and~\ref{algorithm:fragmentgrow}).

\vspace*{1.5mm}
\noindent
\textbf{RQ2 (Scalability).} 
We answer this research question by analysing how increasing the number of parameters in a pDTMC (i.e., the number of transition probabilities specified as rational functions), affects the execution time and complexity of formulae computed by \approach, Storm and PRISM. 
We used the four PL system variants and varied the number of parameters in these models as a percentage of the total model parameters, setting the remaining parameters to randomly selected constant values that preserve the pDTMC structure. 
For all pDTMCs, we started from 1\% (P01), continuing in increments of 10\% (P10), 20\% (P20) etc. until all transition probabilities are given by rational functions (P100).

Table~\ref{Table:PLmodel} shows the PMC results for the four PL variants. 
For all PMC approaches, as expected, increasing the number of pDTMC model parameters incurs a corresponding increase in the execution time and formulae complexity. 
Both Storm and PRISM, however, exhibit exponential growth in these two metrics until P60, and fail to produce any formula from P70 thereafter.
In fact, some of the formulae comprise as many as 3.91M and 1.17M arithmetic operations for PRISM and Storm, respectively. 
In contrast, \approach\ terminates successfully in all cases, consuming at most 94s, indicating
that \approach\ is barely influenced by the increase in model parameters.
This insight is also supported by the result that the \approach-computed closed-form analytical models contain at most 261K operations. 

Considering these results, we have strong empirical evidence indicating that the higher the percentage of transition probabilities specified as rational functions over the total number of transition probabilities, the more apparent is the performance gap favouring \approach\ over Storm and PRISM.
Accordingly, \approach\ can support the derivation of closed-form analytical models for reachability properties whose associated pDTMCs specify as much as 100\% transition probabilities as rational functions in the evaluated pDTMCs.

\vspace*{1.5mm}
\noindent
\textbf{RQ3 (Configurability).} 
Columns \approach\ and \approach\ ($\alpha\!\!=\!\!\infty$) in Table~\ref{Table:FXmodel} compare the \approach\ performance with the soft upper bound for the fragment size enabled and disabled, respectively.
Clearly, using $\alpha$ extends the applicability of the \approach, reduces its execution time and produces analytical models with a smaller number of operations.
Fig.~\ref{fig:Sensitivity} shows the execution time and number of operations for the closed-form analytical models for threshold $\alpha \!\!\in\!\! \{1, .., 29\}$ for the FX system from our running example. 
The general pattern indicates that larger $\alpha$ values, contribute to longer execution times and larger formulae, thus negatively influencing the \approach\ performance.  
Since for larger $\alpha$ the model restructuring techniques are invoked less often, the computed formulae are unsurprisingly more complex.
However, we observe a `sweet spot' for $\alpha \!\in\! \{\!3,\!..,\!14\}$ signifying that fragments of those sizes produce the most effective pDTMC fragmentation. 
These findings illustrate that enhancing \approach\ with mechanisms enabling the automated selection of  $\alpha$ values (e.g., based on the pDTMC structure) can extend further its applicability; this is left for future work.

\begin{figure}
    \centering
    \includegraphics[width=\hsize]{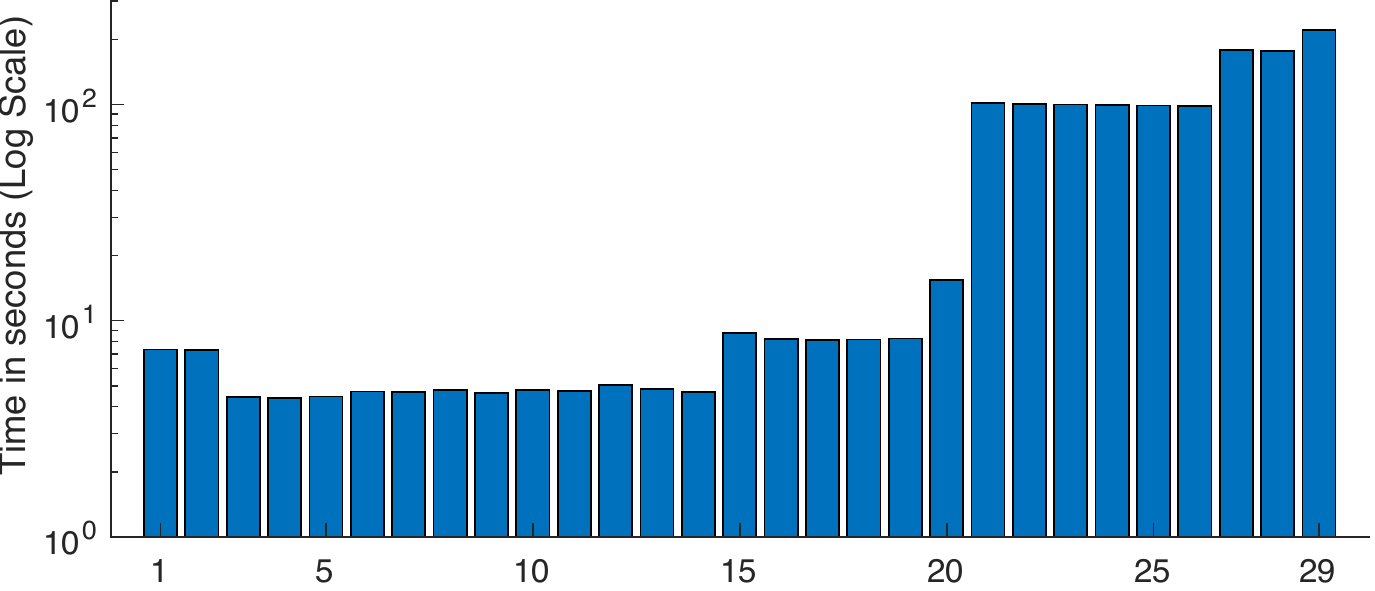}
    
    \vspace*{2.5mm}
    \includegraphics[width=\hsize]{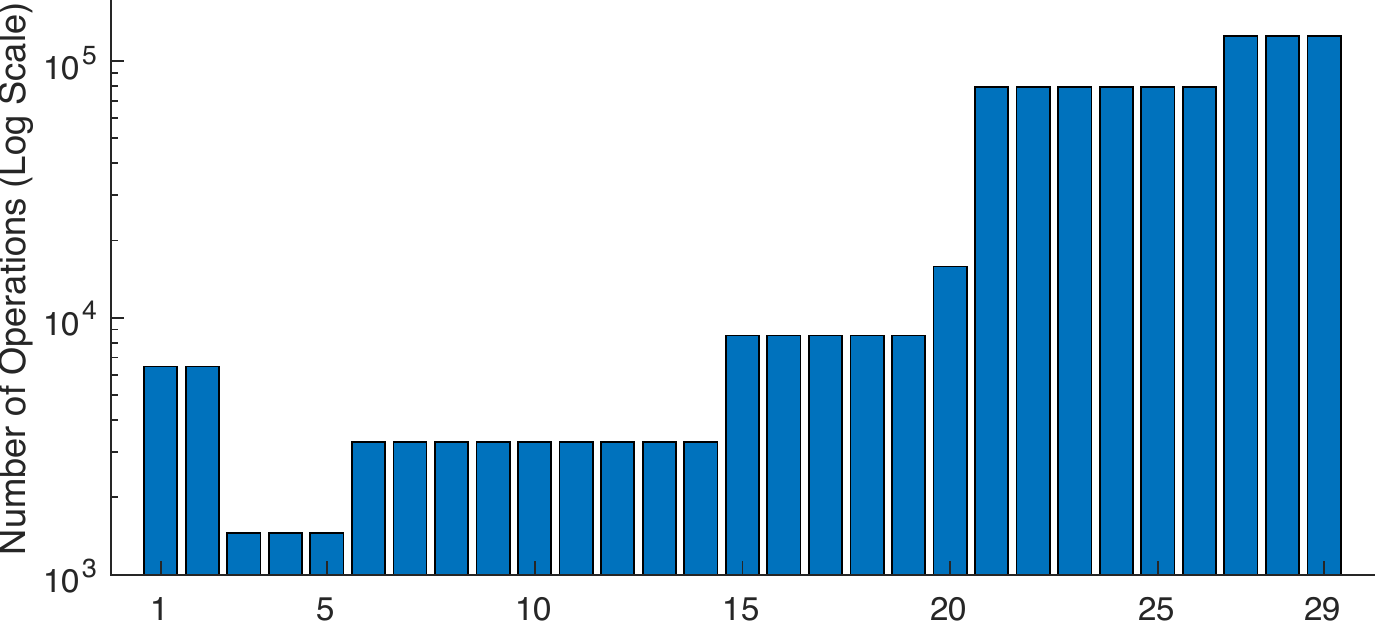}
    \caption{Impact of the threshold $\alpha$ on the \approach\ time (top) and \approach\ expression complexity (bottom) for the FX system}
    \label{fig:Sensitivity}
\end{figure}

\subsection{Threats to Validity}\label{ssec:threats}

We limit \textbf{construct validity threats} that may originate from simplifications and assumptions made when establishing the experimental methodology using the pDTMCs and reachability properties from three publicly-available software systems and processes, also used in related research~\cite{RaduePMC,ghezzi2013model,hajnal2019data,classen2010model,Gerasimou2015:ASE}. 

We mitigate \textbf{internal validity threats} that could introduce bias when identifying cause-effect relationships in our experiments by designing independent research questions and evaluating the effectiveness, scalability and configurability (sensitivity) of \approach. 
To further reduce the risk of biased results, we compared \approach\ against the latest versions of the leading probabilistic model checkers Storm~\cite{storm} and PRISM~\cite{prism} that were available when we conducted the evaluation.
We also performed experiments using multiple variants of the studied software systems and process, and evaluated the correctness of all the \approach-computed formulae by adopting the approach described in Section~\ref{ssec:fpmcapplication}. 
Finally, we enable replication and verification of our findings by making all experimental results publicly available online. 

We limit \textbf{external validity threats} that could affect the generalisability of our findings
by evaluating \approach\ using pDTMCS of systems and processes from three different application domains (i.e., service-based systems~\cite{ameller2016survey,sun2010modeling}, software product lines~\cite{clements2002software,apel2013software}, communication processes~\cite{dressler2010survey,dressler2010bio}).  
Furthermore, we carried out experiments to check that \approach\ can work with pDTMCs containing up to 83 parameters, incurring modest increase in execution time with the computed closed-form analytical models containing significantly fewer arithmetic operations than those produced by the model checkers PRISM and Storm. 
We reduce further the risk that \approach\ might be difficult to use in practice by developing an \approach\ prototype tool that requires the same domain expertise for specifying pDTMCs as for using PRISM and Storm.
However, additional experiments are needed to confirm that \approach\ 
can analyse pDTMCs modelling other software systems and processes than those employed in our evaluation.

\section{Related Work}\label{sec:relatedWork}

Introduced by Daws~\cite{Daws:2004:SPM:2102873.2102899} less than two decades ago, parametric model checking has been significantly advanced through the use of rational PMC expression characteristics such as symmetry and cancellation properties~\cite{Hahn2011}, and of polynomial factorisation and strongly connected component decomposition~\cite{Jansen2014}. Both the former advance (implemented by the model checkers PARAM~\cite{param}, PRISM~\cite{prism} and Storm~\cite{storm}) and the latter (implemented, to the best of our knowledge, only by Storm~\cite{storm}) have greatly improved the scalability of PMC by ensuring that simpler expressions than those from~\cite{Daws:2004:SPM:2102873.2102899} are produced during the parametric model checking process.

However, the objective of all these PMC techniques is to compute a single closed-form expression. The size and complexity of this expression grow rapidly as the analysed pDTMCs become larger and have more parameters, until its computation is no longer feasible, or the expression is so large that its evaluation is impractical. Our \approach\ technique employs a divide-and-conquer strategy that is completely different from the PMC techniques from~\cite{Daws:2004:SPM:2102873.2102899,Hahn2011,Jansen2014}. As such, for many large or complex models, and for models with many parameters, \approach\ replaces the often unfeasible task of computing extremely large expressions with feasible tasks that involve the computation of multiple much simpler expressions. By additionally leveraging the existing PMC techniques for the computation of these simpler expressions, \approach\ manages to successfully handle much larger pDTMCs than previously possible.

Our PMC theoretical framework from~\cite{RaduePMC}, which \approach\ extends and automates, is complementary to the work presented in this paper. As detailed in \S\ref{subsec:fragmentationTheory}, this approach requires manual assembly of the analysed pDTMC by an expert, which is time consuming, error prone, and only feasible for specific classes of component-based systems. To the best of our knowledge, \approach\ is the first approach that can tackle the PMC of parametric Markov models of the complexity and with the number of parameters illustrated in \S\ref{sec:evaluation}.

\section{Conclusion}\label{sec:conclusions}

We presented, implemented and evaluated \approach, an automated model fragmentation technique that enables the fast parametric model checking of reachability properties for systems with complex stochastic behaviour and large numbers of parameters. fPMC complements, leverages and extends the applicability of current PMC techniques and tools, so that closed-form analytical models can be computed for additional classes of systems. Given the usefulness of such analytical models in software engineering and other disciplines, we plan to continue to develop \approach. 

Our next steps in this project will focus on extending the types of properties handled by \approach\ with unbounded until (\S\ref{pmc}) and reward~\cite{andova2003discrete} PCTL properties, 
on developing additional model restructuring techniques to support the \approach\ model fragmentation, and on investigating how the selection of node $z_0$ affects the performance of fPMC, thus improving model fragmentation further via $z_0$-informed selection strategies.
Finally, we will extend fPMC with support for hierarchical fragmentation
so that large fragments and abstract models are further partitioned into smaller models, making PMC applicable to even larger systems.

\section*{Acknowledgement}

This project has received funding from the ORCA Hub PRF project `Continual Verification and Assurance of Robotic Systems under Uncertainty', the UKRI project EP/V026747/1 `Trustworthy Autonomous Systems Node in Resilience', and the Assuring Autonomy International Programme.

\bibliographystyle{./IEEEtran}
\bibliography{reference}


\end{document}